%% file: emission_height.tex
\documentclass[twocolumn]{aastex631}

\usepackage{amsmath}
\usepackage{xcolor}
\usepackage{comment}
\usepackage{csquotes}
\usepackage{{booktabs}}

\newcommand{\discminer}{\textsc{discminer} }
\newcommand{\disksurf}{\texttt{disksurf} }

\received{November 28, 2024}
\revised{February 24, 2025}
\accepted{March 13, 2025}

\begin{document}

\title{exoALMA XV: Interpreting the height of CO emission layer}

\author[0000-0003-4853-5736]{Giovanni P. Rosotti}
\affiliation{Dipartimento di Fisica, Università degli Studi di Milano, Via Celoria 16, 20133 Milano, Italy}

\correspondingauthor{Giovanni P. Rosotti}
\email{giovanni.rosotti@unimi.it}

\author[0000-0003-4663-0318]{Cristiano Longarini}
\affiliation{ Institute of Astronomy, University of Cambridge, Madingley Rd, CB30HA, Cambridge, UK}
\affiliation{Dipartimento di Fisica, Università degli Studi di Milano, Via Celoria 16, 20133 Milano, Italy}

\author[0000-0002-4044-8016]{Teresa Paneque-Carreño}
 \affiliation{European Southern Observatory, Karl-Schwarzschild-Str. 2, D-85748 Garching bei M\"unchen, Germany}
\affiliation{Leiden Observatory, Leiden University, P.O. Box 9513, NL-2300 RA Leiden, The Netherlands}
\affiliation{Department of Astronomy, University of Michigan, 1085 South University Avenue, Ann Arbor, MI 48109, USA}

 \author[0000-0002-2700-9676]{Gianni Cataldi} 
\affiliation{National Astronomical Observatory of Japan, 2-21-1 Osawa, Mitaka, Tokyo 181-8588, Japan}

\author[0000-0002-5503-5476]{Maria Galloway-Sprietsma}
\affiliation{Department of Astronomy, University of Florida, Gainesville, FL 32611, USA}


\author[0000-0003-2253-2270]{Sean M. Andrews}
\affiliation{Center for Astrophysics | Harvard \& Smithsonian, Cambridge, MA 02138, USA}

\author[0000-0001-7258-770X]{Jaehan Bae}
\affiliation{Department of Astronomy, University of Florida, Gainesville, FL 32611, USA}

\author[0000-0001-6378-7873]{Marcelo Barraza-Alfaro}
\affiliation{Department of Earth, Atmospheric, and Planetary Sciences, Massachusetts Institute of Technology, Cambridge, MA 02139, USA}

 \author[0000-0002-7695-7605]{Myriam Benisty}
\affiliation{Université Côte d’Azur, Observatoire de la Côte d’Azur, CNRS, Laboratoire Lagrange, 06304 Nice, France}
 \affiliation{Max-Planck Institute for Astronomy (MPIA), Königstuhl 17, 69117 Heidelberg, Germany}

\author[0000-0003-2045-2154]{Pietro Curone} 
\affiliation{Dipartimento di Fisica, Università degli Studi di Milano, Via Celoria 16, 20133 Milano, Italy}
\affiliation{Departamento de Astronom\'ia, Universidad de Chile, Camino El Observatorio 1515, Las Condes, Santiago, Chile}

\author[0000-0002-1483-8811]{Ian Czekala}
\affiliation{School of Physics \& Astronomy, University of St. Andrews, North Haugh, St. Andrews KY16 9SS, UK}

 \author[0000-0003-4689-2684]{Stefano Facchini}
\affiliation{Dipartimento di Fisica, Università degli Studi di Milano, Via Celoria 16, 20133 Milano, Italy}

\author[0000-0003-4679-4072]{Daniele Fasano} 
\affiliation{Université Côte d’Azur, Observatoire de la Côte d’Azur, CNRS, Laboratoire Lagrange, 06304 Nice, France}

\author[0000-0002-9298-3029]{Mario Flock} 
\affiliation{Max-Planck Institute for Astronomy (MPIA), Königstuhl 17, 69117 Heidelberg, Germany}

\author[0000-0003-1117-9213]{Misato Fukagawa} 
\affiliation{National Astronomical Observatory of Japan, 2-21-1 Osawa, Mitaka, Tokyo 181-8588, Japan}

\author[0000-0002-5910-4598]{Himanshi Garg}
\affiliation{School of Physics and Astronomy, Monash University, Clayton VIC 3800, Australia}

 \author[0000-0002-8138-0425]{Cassandra Hall} 
\affiliation{Department of Physics and Astronomy, The University of Georgia, Athens, GA 30602, USA}
\affiliation{Center for Simulational Physics, The University of Georgia, Athens, GA 30602, USA}
\affiliation{Institute for Artificial Intelligence, The University of Georgia, Athens, GA, 30602, USA}

\author[0000-0001-6947-6072]{Jane Huang} 
\affiliation{Department of Astronomy, Columbia University, 538 W. 120th Street, Pupin Hall, New York, NY, USA}

\author[0000-0003-1008-1142]{John~D.~Ilee} 
\affiliation{School of Physics and Astronomy, University of Leeds, Leeds, UK, LS2 9JT}

 \author[0000-0001-8446-3026]{Andr\'es F. Izquierdo} 
  \altaffiliation{NASA Hubble Fellowship Program Sagan Fellow}
 \affiliation{Department of Astronomy, University of Florida, Gainesville, FL 32611, USA}
 \affiliation{Leiden Observatory, Leiden University, P.O. Box 9513, NL-2300 RA Leiden, The Netherlands}
 \affiliation{European Southern Observatory, Karl-Schwarzschild-Str. 2, D-85748 Garching bei M\"unchen, Germany}

\author[0000-0001-7235-2417]{Kazuhiro Kanagawa} 
\affiliation{College of Science, Ibaraki University, 2-1-1 Bunkyo, Mito, Ibaraki 310-8512, Japan}

\author[0000-0002-8896-9435]{Geoffroy Lesur} 
\affiliation{Univ. Grenoble Alpes, CNRS, IPAG, 38000 Grenoble, France}


\author[0000-0002-2357-7692]{Giuseppe Lodato} 
\affiliation{Dipartimento di Fisica, Università degli Studi di Milano, Via Celoria 16, 20133 Milano, Italy}

\author[0000-0002-8932-1219]{Ryan A. Loomis}
\affiliation{National Radio Astronomy Observatory, 520 Edgemont Rd., Charlottesville, VA 22903, USA}


\author[0000-0003-4039-8933]{Ryuta Orihara} 
\affiliation{College of Science, Ibaraki University, 2-1-1 Bunkyo, Mito, Ibaraki 310-8512, Japan}

\author[0000-0001-5907-5179]{Christophe Pinte}
\affiliation{Univ. Grenoble Alpes, CNRS, IPAG, 38000 Grenoble, France}
\affiliation{School of Physics and Astronomy, Monash University, Clayton VIC 3800, Australia}

\author[0000-0002-4716-4235]{Daniel J. Price} 
\affiliation{School of Physics and Astronomy, Monash University, Clayton VIC 3800, Australia}


\author[0000-0002-0491-143X]{Jochen Stadler} 
 \affiliation{Université Côte d’Azur, Observatoire de la Côte d’Azur, CNRS, Laboratoire Lagrange, 06304 Nice, France}

 \author[0000-0003-1534-5186]{Richard Teague}
 \affiliation{Department of Earth, Atmospheric, and Planetary Sciences, Massachusetts Institute of Technology, Cambridge, MA 02139, USA}

\author[0000-0002-3468-9577]{Gaylor Wafflard-Fernandez} 
\affiliation{Univ. Grenoble Alpes, CNRS, IPAG, 38000 Grenoble, France}

\author[0000-0002-7501-9801]{Andrew J. Winter}
\affiliation{Université Côte d’Azur, Observatoire de la Côte d’Azur, CNRS, Laboratoire Lagrange, 06304 Nice, France}
\affiliation{Max-Planck Institute for Astronomy (MPIA), Königstuhl 17, 69117 Heidelberg, Germany}

\author[0000-0002-7212-2416]{Lisa W\"olfer} 
\affiliation{Department of Earth, Atmospheric, and Planetary Sciences, Massachusetts Institute of Technology, Cambridge, MA 02139, USA}

\author[0000-0003-1412-893X]{Hsi-Wei Yen} 
\affiliation{Academia Sinica Institute of Astronomy \& Astrophysics, 11F of Astronomy-Mathematics Building, AS/NTU, No.1, Sec. 4, Roosevelt Rd, Taipei 10617, Taiwan}

\author[0000-0001-8002-8473	]{Tomohiro C. Yoshida} 
\affiliation{National Astronomical Observatory of Japan, 2-21-1 Osawa, Mitaka, Tokyo 181-8588, Japan}
\affiliation{Department of Astronomical Science, The Graduate University for Advanced Studies, SOKENDAI, 2-21-1 Osawa, Mitaka, Tokyo 181-8588, Japan}

\author[0000-0001-9319-1296	]{Brianna Zawadzki} 
\affiliation{Department of Astronomy, Van Vleck Observatory, Wesleyan University, 96 Foss Hill Drive, Middletown, CT 06459, USA}
\affiliation{Department of Astronomy \& Astrophysics, 525 Davey Laboratory, The Pennsylvania State University, University Park, PA 16802, USA}



\begin{abstract}

The availability of exquisite data and the development of new analysis techniques have enabled the study of emitting heights in proto-planetary disks. In this paper we introduce a simple model linking the emitting height of CO to the disk surface density and temperature structure. We then apply the model to measurements of the emitting height and disk temperature conducted as part of exoALMA, integrated with additional legacy measurements from the MAPS Large Programme, to derive CO column densities and surface density profiles (assuming a CO abundance) for a total of 14 disks. A unique feature of the method we introduce to measure surface densities is that it can be applied to optically thick observations, rather than optically thin as conventionally done. While we use our method on a sample of well studied disks where temperature structures have been derived using two emission lines, we show that reasonably accurate estimates can be obtained also when only one molecular transition is available. With our method we obtain independent constraints from $^{12}$CO and $^{13}$CO and we find they are in general good agreement using the standard $^{12}$C/$^{13}$C isotopic ratio. The masses derived from our method are systematically lower compared with the values derived dynamically from the rotation curve if using an ISM CO abundance, implying that CO is depleted by a median factor $\sim$20 with respect to the ISM value, in line with other works that find that CO is depleted in proto-planetary disks.

\end{abstract}



\section{Introduction} \label{sec:intro}
In the last few years the telescope ALMA has rapidly transformed the observational field of proto-planetary disks. Thanks to the combination of high spatial and spectral resolution and the high sensitivity, ALMA has allowed us to study in detail disk kinematics (see \citealt{PinteReview} for a review).  In parallel, new analysis tools have been developed \citep{TeagueEddy,CasassusPerez2019,DiscMinerPaper} to make the most use of the data.

One application of spatially and spectrally resolved observations is to measure the height of the emission surface for molecular lines \citep{Rosenfeld2013,Pinte2018}, and the temperature at the emitting surface since the emission is optically thick. These kinds of measurements are now available for a growing sample of disks \citep{Law2021,Law2022,Law2023} and also for multiple molecules \citep[e.g.,][]{Paneque-Carreno2023,Urbina2024}, while initial efforts focused mostly on the CO molecule on account of its brightness.

This paper is part of the exoALMA Large Programme \citep{exoALMA:Richard}, which consists in observations designed to perform kinematic analysis targeting 15 proto-planetary disks. As part of a dedicated paper series, \citet{exoALMA:Andres} and \citet{exoALMA:Maria} derive emitting heights and temperatures for the disks of the sample with two different techniques. In this work, we adopt the emitting heights derived parametrically by \citet{exoALMA:Andres} and the temperatures derived by \citet{exoALMA:Maria}.

While these measurements are becoming routine, comparatively little effort has been invested to investigate what sets the emitting height and what the measurements tell us about the disk structure and composition. Addressing this issue is the focus of this paper. In particular, we aim to develop a model to quantitatively link a given density and temperature structure with the emitting height. Because the observations put constraints on both the latter and the temperature structure, our aim is then to invert the problem and use the observational constraints to measure the disk density.

The paper is structured as follows. In section \ref{sec:model}, assuming hydrostatic equilibrium in the vertical direction, we construct an analytical model to link the emitting height with the disk surface density, with assumptions on the disk vertical temperature profile. The model depends on a free parameter describing photo-dissociation, which we calibrate against thermo-chemical models in section \ref{sec:calibration}. We then apply the model to the exoALMA sample (integrated with additional disks from MAPS) in section \ref{sec:results}, deriving surface densities for our targets. We discuss the implications of our results in section \ref{sec:discussion}, in particular with respect to the issues of CO abundance and isotopic ratio, and we finally draw our conclusions in \ref{sec:conclusions}.

\section{Model for emission height}
\label{sec:model}

In this work we use the Eddington-Barbier approximation (\citealt{1943AnAp....6..113B}, but see \citealt{2018OAst...27...76P} for an interesting historical perspective) to make the assumption that the CO emission height traces the location where $\tau \simeq 2/3$, where $\tau$ is measured from the observer to the emitting layer. We call $N_\tau$ the column density of CO for which this happens. Unless the disk is viewed nearly edge-on, and using the fact that the disk is vertically thin, the line of sight can be approximated as almost vertical in the disk frame, and therefore probes a single radius. We thus need to estimate the optical depth of the emission line along vertical rays to interpret the observed heights. 

In addition, we also need to take into account that in the upper layers of the disk molecules are photo-dissociated (e.g., \citealt{Aikawa2002}) by the UV radiation (either emitted from the star, or from the interstellar field). This implies that the uppermost layers of the disk do not provide any contribution to the line optical depth, since they contain virtually no CO. Thermochemical models find that the CO abundance rises quickly over a vertically narrow region in which CO self-shields effectively from UV radiation; therefore, it is a good approximation, and often employed \citep[e.g.,][]{Qi2011,WilliamsBest,Flaherty2015,Toci2021}, to assume that there is no contribution to the line optical depth up to some height $z_{ph}$, the photo-dissociation layer, while for $z<z_{ph}$ CO has a constant abundance $x_\mathrm{CO}$. While precise constraints on photodissociation would require detailed thermochemistry to be followed exactly, in this work we make the simplifying assumption that a fixed vertical column of the molecule $N_\mathrm{ph}$ is required to self-shield from UV radiation, consistent with what is found in more complex calculations (e.g. \citealt{vanDishoeckBlack1988,Visser2009}). This column is typically found to be somewhere around $10^{16} \ \mathrm{cm}^{-2}$, but we will compare with thermo-chemical models in section \ref{sec:calibration} to find the best $N_\mathrm{ph}$ value. Note that, under the assumption that CO self-shields, in this paper we define this quantity as the equivalent column of CO required to self-shield, and not as the \textit{total} gas column - this is because the latter would depend on CO abundance, whereas $N_\mathrm{ph}$ we employ does not. Combining the two requirements, we expect the emission layer to be at a height where the column density has a critical value $N_\mathrm{crit}=N_\mathrm{ph}+N_\tau$. We now focus in section \ref{sec:opacity} on the calculation of $N_\tau$ and then discuss in the rest of section \ref{sec:model} our model for the emitting height and how to turn measurements of the emitting height into measurements of the surface density, taking into account that disks have a vertical density and temperature stratification. We will leave $N_\mathrm{ph}$ as a free parameter in this section and discuss the procedure we followed to set its value in section \ref{sec:calibration}.

\subsection{Line opacity and critical column probed by the emitting layer}
\label{sec:opacity}

We recap here standard radiative transfer formulae (see e.g. \citealt{2016era..book.....C,Radex}). Under the local thermodynamical equilibrium (LTE) condition, which thermochemical models show to hold true for low J transitions of the CO molecule, the opacity coefficient\footnote{We follow here the notation of \citet{2016era..book.....C} in which $\tau=\int \kappa \mathrm{d}s$. Note that in this notation the symbol $\kappa$ is not a \textit{mass} opacity.} at a frequency $\nu$ of a line with emission frequency $\nu_{ul}$, corresponding to the transition from an upper level with quantum degeneracy $g_u$ to a lower level with quantum degeneracy $g_l$, is given by
\begin{equation}
    \label{eq:tau}
    \kappa (\nu)= \frac{c^2}{8\pi \nu_{ul}^2} A_{ul} \frac{g_u}{g_l} x_l n_{mol}\left[ 1-\exp\left({-\frac{h\nu_{ul}}{k_bT}}\right)\right] \phi(\nu),
\end{equation}
where $A_{ul}$ is the Einstein coefficient of the transition, $x_l$ is the fraction of molecules in the lower level of the transition, $n_{mol}$ is the volume density of the molecule, $k_b$ the Boltzmann constant, and $\phi(\nu)$ is the line profile. Because we are in LTE, the fraction of molecules in a given energy level is
\begin{equation}
    x_i = \frac{g_i }{Z(T)}\exp\left({-\frac{E_i}{k_b T}}\right),
\end{equation}
where $E_i$ is the energy of the level and $Z (T)$ the partition function. We estimate the partition function using the expansion $Z=1/3 + k_b T/hB_0$, where $B_0$ is the rotational constant of the molecule. We verified that there is no measurable difference in our results from explicitly calculating the partition function by summing over all the rotational energy levels in the Leiden molecular database \citep{LeidenMolecularDatabase} for the range of temperatures we are interested in ($T>10$ K).

The last factor to discuss is the line profile. Lines in proto-planetary disks are significantly Doppler-shifted by Keplerian rotation, but here we are interested in the optical depth at a specific point of the disk. We can therefore neglect rotation, as long as the velocity does not change significantly along the line of sight. While disks do have \textit{vertical} velocity gradients \citep{TakeuchiLin2002,Rosotti2020,Martire2024}, these are typically small (e.g., much smaller than \textit{radial} velocity gradients), justifying our assumption. We also note that in this work we are interested in estimating the height of the emitting layer, and not the total optical depth of the line (where the Doppler shift from the upper layers to the midplane, or to the back surface, can be significant). In practice, since disks have a strong vertical density gradient as a result of hydrostatic equilibrium, most of the optical depth of the line is generated in the proximity of the emitting layer. A corollary of this fact is that the integrations we will write in the rest of this section, although formally over the whole line of sight up to to the emitting layer, are dominated by the region close to the emission layer that we probe observationally. Assuming no Doppler shift across the spatial length-scale of the layer is therefore a good assumption. 

We also neglect turbulent broadening, since even in cases where this has been measured \citep{FlahertyDMTau,FlahertyIMLup,Paneque-CarrenoIMLup} it is at most comparable to the thermal value, and smaller in all other cases (see \citealt{Rosotti2023} for a summary). Finally, although pressure broadening has been detected in TW Hya \citep{Yoshida2022} and \citet{Yoshida_exoALMA} report another detection as part of this paper series, pressure broadening is a discernible contribution only in the wings of the line profile coming from the disk midplane, while here we are concerned with the line center originating in a location where the density is significantly lower. Therefore, in what follows we only consider thermal broadening; this translates to a Gaussian line profile, which at the line center evaluates to:
\begin{equation}
    \left. \phi(\nu) \right|_{\nu_{ul}} = \frac{c}{\nu_{ul}} \frac{1}{\sqrt{\pi} v_D},
\end{equation}
so that the line profile is normalized to 1. The thermal line-width $v_D$ is
\begin{equation}
    v_D = \sqrt{2 k_b T / m_{mol}},
\end{equation}
where $m_{mol}$ is the mass of the molecule. To see more in detail why we can neglect turbulent line broadening, consider that the turbulent line-width would add in quadrature to thermal line-width. Since the opacity (\autoref{eq:tau}) is inversely proportional to the line-width, only a turbulence width much larger than the thermal width would significantly change the results we will now discuss.

\begin{figure}
    \centering
    \includegraphics[width=\columnwidth]{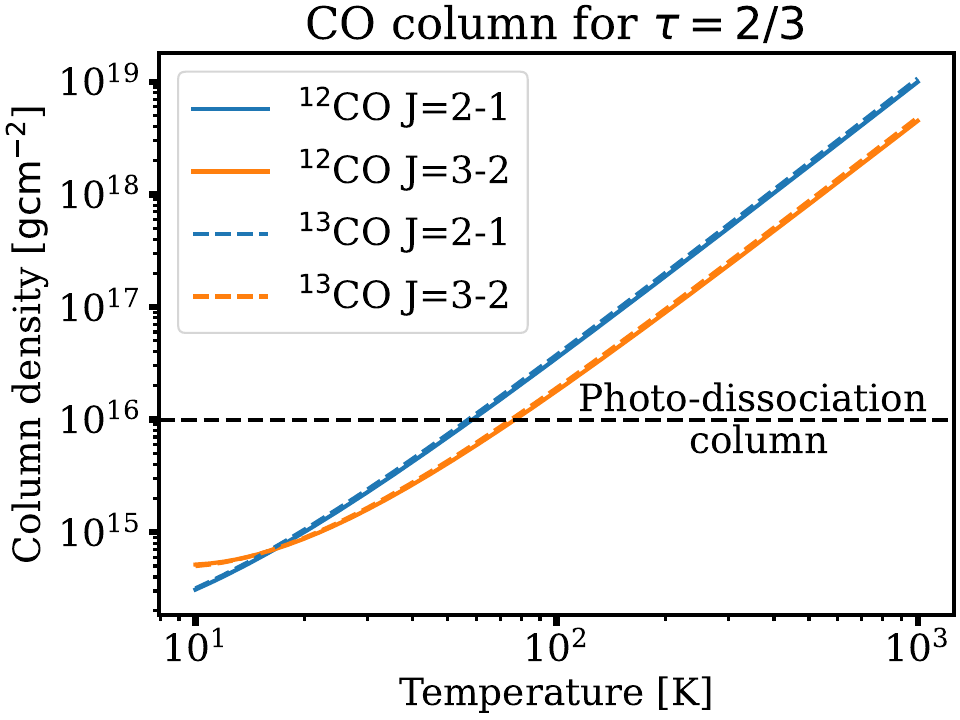}
    \caption{$N_\tau$, the CO column to have $\tau=2/3$ for the J=3-2 and J=2-1 transition of $^{12}$CO and $^{13}$CO, as a function of temperature. The photo-dissociation column value we show is only a reference and it will be estimated more precisely in section \ref{sec:calibration}.}
    \label{fig:critical_column}
\end{figure}

Although these are textbook relations, to illustrate what they imply in the context of proto-planetary disks, we show\footnote{We benchmarked our calculation against RADEX \citep{Radex} and confirm it yields the same results, except for a factor $\sim$1.06. This is known and descends from the fact that RADEX assumes a rectangular, rather than Gaussian, line profile.} in \autoref{fig:critical_column} $N_\tau$, the column at which $\tau=2/3$, assuming an isothermal slab of material. To compute this we use that the optical depth is $\tau=\int \kappa (T) \mathrm{d} s$,  where $\kappa(T)$ is given by \autoref{eq:tau}. In the isothermal slab the expression can be rewritten as $ \sigma(T) \int n_{mol} \mathrm{d} s = \sigma(T) N_{mol}$, where we gathered all the terms except the density of the molecule in a factor $\sigma(T)$ - dimensionally, this factor is an area and can be regarded as the cross section to radiation. This highlights that for an isothermal slab the optical depth depends only on the column of material. Because of the transitions we will later use in this paper, we show the results for both the J=3-2 and J=2-1 transitions for $^{12}$CO and $^{13}$CO. There is very little difference between the two isotopologues, while the difference is more substantial for the two transitions, but we still expect them to trace roughly the same region. We note that $N_\tau$ is a strong function of temperature, and accurate estimates of the temperature are therefore critical in order to estimate it correctly. This dependence is driven by several factors in \autoref{eq:tau} that depend on temperature, namely thermal broadening, the partition function, and the Boltzmann factor appearing in the fractional occupancy of the levels. In general, for these low J transitions all these factors tend to \textit{decrease} the line opacity at higher temperatures, since the molecules are spread over a wide velocity range and more quantum levels. In addition, stimulated emission (the negative exponential factor) further reduces the net opacity, since its relevance is proportional to the difference in occupation fraction between upper and lower levels, but this becomes progressively smaller at high temperatures. This drives the general trend that the critical column \textit{increases} with temperature. Because disks have a radial (as well as vertical) temperature gradient, this implies that the emitting layer does \textit{not} trace a fixed column of material, but it traces different columns of material at different radii. Finally, $N_\tau$ is of the same order as $N_{ph}$ (at least for low temperatures), implying that we need to accurately estimate $N_{ph}$ for the model to be accurate.

\subsection{The \enquote{direct} problem: computing the emitting height given a surface density profile}
Armed with \autoref{eq:tau}, we are now equipped to solve what we call the \enquote{direct} problem, which consists of determining the emitting height given a surface density profile $\Sigma(R)$. This is normally done through radiative transfer models, either using simplified assumptions for photo-dissociation \citep[e.g.,][]{WilliamsBest,Pinte2018,Toci2021} or consistent thermochemistry \citep[e.g.,][]{Bruderer2013,Woitke2016}. To do this, we first need an expression for the vertical density profile $\rho(z)$. In the commonly used vertically isothermal case, this is simply
\begin{equation}
\rho = \rho_0 \exp \left( -\frac{z^2}{2 H^2} \right),
\end{equation}
where $H=c_s/\Omega$ is the disk scale-height and $\rho_0=\Sigma/\sqrt{2\pi} H$ is the density at the disk midplane. In the more general vertically stratified case, the vertical density profile is given by \citep[e.g.,][]{Rosotti2020HD100453,Martire2024}:
\begin{equation}
    \rho (z)=  \rho_0 \frac{c_\mathrm{s,midplane}^2}{c_s^2(z)} \exp \left( - \int_0^z \frac{\Omega_k^2(z') z' \mathrm{d}z'}{c_s^2(z')} \right),
    \label{eq:vertical_rho}
\end{equation}
where $c_s(z)$ is the vertical profile of the sound speed and $\Omega_k^2(z)=GM_\ast / (r^2+z^2)^{3/2}$ is the Keplerian frequency at a height $z$. Because most of the mass is at the disk midplane, we find (see also \citealt{Flock2013}) that for reasonable choices of the vertical temperature profile $\rho_0=\Sigma/\sqrt{2\pi} H$ still holds very well (with a typical accuracy of a few per cent), as long as the scale-height is estimated using the temperature value at the disk midplane. In practice however we integrate \autoref{eq:vertical_rho} to link $\rho_0$ with $\Sigma$.

Once the vertical density profile is known, we can find the photo-dissociation height $z_\mathrm{ph}$, which is given implicitly by
\begin{equation}
\label{eq:ph}
    -\int_{+\infty}^{z_\mathrm{ph}} n_\mathrm{CO}(z) \mathrm{d}z = N_\mathrm{ph},
\end{equation}
where $n_\mathrm{CO}$ is the number density of CO molecules. To link it with the mass density $\rho$, we introduce a factor $x_\mathrm{CO}$ to parametrize the CO molecular abundance in the disk; with this substitution,
\begin{equation}
n_\mathrm{CO}= \frac{\rho x_\mathrm{CO}}{\mu m_h},
\end{equation}
where $\mu=2.35$ is the mean molecular weight. We will discuss the value of $x_\mathrm{CO}$ in section \ref{sec:co_abundance_model}. At this point we can finally integrate the equation of radiative transfer to find the height $z_\mathrm{em}$ where the line reaches an optical depth of two thirds:
\begin{equation}
\label{eq:z_em_implicit}
    -\int_{z_\mathrm{ph}}^{z_\mathrm{em}} \kappa(n_\mathrm{CO},T (z))  \frac{\mathrm{d}z}{\cos i} = 2/3,
\end{equation}
where $\kappa(T (z))$ is the line opacity we derived in section \ref{sec:opacity}. This is a function of vertical height since it depends on the density of the CO molecule, and a further dependence is introduced by the dependence of the temperature with the vertical coordinate. The factor $\cos i$ (with $i$ the disk inclination) accounts for the inclined line of sight to the observer. To highlight more the dependence on the vertical coordinate, we can also rewrite this as
\begin{equation}
\label{eq:z_em_implicit_2}
        -\int_{z_\mathrm{ph}}^{z_\mathrm{em}} \sigma(T (z)) n_\mathrm{CO} (z)  \frac{\mathrm{d}z}{\cos i} = 2/3,
\end{equation}
where $\sigma(T)$ is the cross section we introduced in section \ref{sec:opacity}.

For the vertically stratified case, these integrals, and therefore the estimation of the emitting layer, can be computed numerically. For the isothermal case, we can do a further step and derive an analytical solution; \autoref{eq:z_em_implicit} evaluates to
\begin{equation}
\label{eq:z_12}
\begin{split}
N_\mathrm{crit,12}=\int_{z_\mathrm{em,12}}^\infty n_\mathrm{CO} (z) \mathrm{d}z \\ =
\int_{z_\mathrm{em,12}}^\infty \frac{ x_\mathrm{CO}}{\mu m_h} \rho_0 \exp \left( -\frac{z^2}{2 H^2} \right) \mathrm{d} z  \\ =
\frac{ x_\mathrm{CO} \rho_0}{\mu m_h} \sqrt{\frac{\pi}{2}} H \mathrm{erfc} \left( \frac{z_\mathrm{em,12}}{\sqrt{2} H} \right),
\end{split}
\end{equation}
where $N_\mathrm{crit,12}=N_\tau \cos i + N_{ph}$ is the critical column for $^{12}$CO and $\mathrm{erfc}$ is the complementary error function, defined as
\begin{equation}
\mathrm{erfc} (x) = \sqrt{\frac{2}{\pi}} \int_x^\infty e^{-t^2} \mathrm{d}t.
\end{equation}
\autoref{eq:z_12} is an implicit equation for the emitting height; the solution can be written formally as
\begin{equation}
z_\mathrm{em,12} = \sqrt{2} H \mathrm{erfc}^{-1} \left( \frac{\mu m_H}{x_\mathrm{CO} \rho_0} \sqrt{\frac{2}{\pi}} \frac{N_\mathrm{crit,12}}{H} \right).
\end{equation}

Since $\mathrm{erfc}^{-1}$ is a decreasing function, the equation shows that, as we may naively expect, the emitting height increases both with the midplane density and with the scale-height $H$. Finally, we note that $\rho H \propto \Sigma$, where $\Sigma$ is the disk surface density, and we thus expect the emitting height to increase also with the surface density, and correspondingly the disk mass.

Regardless of the vertical density profile, \autoref{eq:z_em_implicit} has the self-shielding column $N_{ph}$ and the CO abundance $x_\mathrm{CO}$ as free parameters. We will calibrate the former against thermochemical models in section \ref{sec:calibration} and discuss the latter in section \ref{sec:co_abundance_model}, as well as constrain it from observations in section \ref{sec:co_depletion}.

\subsection{The \enquote{inverse} problem: retrieving the surface density profile given an observed emitting height}
The \enquote{direct} problem we exposed above illustrates the usefulness of our model, since it allows determining the emitting height with simple calculations rather than setting up a full thermochemical calculation, provided that suitable parameterizations for the temperature and CO abundance be available. However, when intepreting observations, one is normally in the opposite situation: the observations provide an emitting height and we are interested in knowing what constraints it gives on the surface density profile of the disk. We call this the \enquote{inverse} problem. We can regard this problem as an extrapolation from the conditions at the emitting height down to the disk midplane. We detail in what follows how we solve the problem, for the isothermal and vertically stratified case.

\subsubsection{The isothermal case}
\label{sec:inverse_isothermal}

We first note that the inversion is relatively trivial for the isothermal case (\autoref{eq:z_12}). This yields:

\begin{equation}
\label{eq:rho0_isothermal}
\rho_0 = \frac{\mu m_H}{x_\mathrm{CO}} \sqrt{\frac{2}{\pi}} \frac{N_\mathrm{crit,12}}{H} \left( \mathrm{erfc} \left( \frac{z_\mathrm{em,12}}{\sqrt{2} H} \right) \right)^{-1}.
\end{equation}

We can write an analogous expression for $^{13}$CO:
\begin{equation}
\label{eq:rho0_isothermal_13}
\rho_0 = \frac{\mu m_H}{x_\mathrm{CO}} \sqrt{\frac{2}{\pi}} \frac{N_\mathrm{crit,13}}{H} \left( \mathrm{erfc} \left( \frac{z_\mathrm{em,13}}{\sqrt{2} H} \right) \right)^{-1},
\end{equation}
where we have assumed that $^{13}$CO is a factor $^{12}$C/$^{13}$C $\simeq 77$ \citep{isotopic_abundance} less abundant and we have allowed for the general possibility that $N_\mathrm{crit,13}$ is different from $N_\mathrm{crit,12}$. Physically speaking, this arises since the two molecules have slightly different parameters for the transition. We will also use this difference to account for the fact that observationally the two molecules have different brightness temperatures\footnote{Clearly this observational fact is in contradiction with the assumption that the disk is vertically isothermal. We will assess in the rest of the paper how well the isothermal assumption holds.}, which leads to different values of $N_\tau$ for the two isotopologues. We note that recently \citet{Paneque-Carreno2023} presented a similar analysis. In that case however the authors assumed a surface density profile and used the constraint provided by the observed emitting height of $^{12}$CO to derive the disk scale-height, while in this work we use the emitting height to set constraints on the surface density profile.

The two expressions can be evaluated provided that we know the scale-height of the disk $H$. Indeed, the scale-height is needed since, as we explained when introducing the \enquote{inverse} problem, we need to extrapolate from the emitting height to the midplane. We thus need more information in the form of independent constraints on the disk temperature. For the isothermal model we discuss two different approaches to retrieve this information. The first consists in using the line brightness temperature under the assumption that the emission is optically thick, which is often used in disks to measure the disk temperature.

To introduce the second approach, we note that \autoref{eq:rho0_isothermal} and \ref{eq:rho0_isothermal_13} constitute a system of two equations and two unknowns ($\rho_0$ and $H$); it is therefore possible to solve for the two unknowns and determine $H$ in this way. To this end, it is useful to take the ratio between \autoref{eq:z_12} and the equivalent for $^{13}$CO, which yields a relatively simple equation for $H$:
\begin{equation}
\label{eq:H_isothermal}
\frac{^{12}\mathrm{C}}{^{13}\mathrm{C}} \frac{N_\mathrm{crit,13}}{N_\mathrm{crit,12}} \mathrm{erfc} \left( \frac{z_\mathrm{em,12}}{\sqrt{2} H} \right) = \mathrm{erfc} \left( \frac{z_\mathrm{em,13}}{\sqrt{2} H} \right)
\end{equation}
that we can apply directly to the data, if we have measurements of both emitting layers. The equation has a straightforward interpretation as it is merely saying that the column at the $^{13}$CO emitting layer must be a factor $\frac{^{12}\mathrm{C}}{^{13}\mathrm{C}} \frac{N_\mathrm{crit,13}}{N_\mathrm{crit,12}}$ higher than the column at the $^{12}$CO emitting layer. Interestingly, all other quantities cancel out, and therefore $H$ depends only on the isotopic ratio. The equation needs to be solved numerically, which can be done using a standard root finding algorithm (we use the Brent solver as implemented in the scipy function brentq). Once $H$ is known, we can apply \autoref{eq:rho0_isothermal} to retrieve the density in the midplane.



\subsubsection{The vertically stratified case}
We first note that this case can be solved only if we know the vertical temperature profile $T(z)$. This is because the temperature potentially takes a different value at every vertical coordinate $z$, and for this reason we can no longer describe the temperature with a single parameter. In the context of exoALMA, the radial and vertical temperature profiles are fitted to the observed brightness temperatures in the work of \citet{exoALMA:Maria}. While we refer to that work for the details of the temperatures we employ, for the remainder of this section we will simply assume the temperature profile is known.

A fixed temperature profile implies, solving the hydrostatic equilibrium equation, that the density profile is also fixed, save for a normalization factor. Determining this factor is our goal. It is convenient to write the vertical density profile as
\begin{equation}
\rho (z) = \rho_0 f(z)
\end{equation}
introducing the value of the density in the midplane $\rho_0$ and encoding the vertical dependency of the density in the function $f(z)$. It is trivial to define along the same lines the CO number density in the midplane $n_0$. With this definition, the photo-dissociation condition (\autoref{eq:ph}) becomes
\begin{equation}
-n_0 \int_{+\infty}^{z_\mathrm{ph}} f(z) \mathrm{d}z = N_\mathrm{ph}
\label{eq:photodissociation}
\end{equation}
while the optical depth condition (\autoref{eq:z_em_implicit}, which we rewrite using the notation of \autoref{eq:z_em_implicit_2}) becomes
\begin{equation}
-n_0 \int_{z_\mathrm{ph}}^{z_\mathrm{em}} \sigma(T(z)) f(z) \frac{\mathrm{d}z}{\cos i} = 2/3.
\end{equation}
For computational reasons we rewrite this condition as
\begin{equation}
-\frac{n_0}{\cos i} \left( \int_{+\infty}^{z_\mathrm{em}} \sigma(T) f(z) \mathrm{d}z - \int_{+\infty}^{z_\mathrm{ph}} \sigma(T) f(z) \mathrm{d}z \right)=2/3.
\label{eq:opticaldepth}
\end{equation}
Practically speaking, the two conditions contain integrals that we can tabulate once for each radius, irrespective of the value of $n_0$; this is thus a problem that can be solved very efficiently, with typical run times shorter than 1 second on a standard desktop CPU to solve the inverse problem for a given disk. Operationally, we use a root finding algorithm (we use the Brent solver as for the isothermal case) to find the root of \autoref{eq:opticaldepth} in terms of $n_0$. This requires first, for a given choice of $n_0$, to solve \autoref{eq:photodissociation} for $z_{ph}$, since it is needed in \autoref{eq:opticaldepth}. Since we tabulate the integral appearing in \autoref{eq:photodissociation}, this can be done efficiently using linear interpolation. Finally, we note that the Brent solver requires specifying a lower and upper limit bracketing the root. In order to do this we initially neglect the photo-dissociation column $N_{ph}$ (i.e. we assume $z_{ph}=\infty$), which provides a reasonable first guess for $\rho_0$ by inverting \autoref{eq:opticaldepth}. We use values a factor 1000 above and below this initial estimate to provide the search range; we have verified this range is sufficiently broad to always encompass the actual root\footnote{As expected, the root is always \textit{higher} than the initial estimate.}.

\subsection{The role of the unknown CO abundance}
\label{sec:co_abundance_model}

We remark that strictly speaking the constraints we derive in this work only concern the CO molecule. Our procedure amounts to turn a measurement of the height at which CO emission is optically thick into a measurement of the total CO column. Only under the assumption that there is a constant scaling factor between the total number density and the CO number density, namely the $x_\mathrm{CO}$ factor, can we turn this constraint into a constraint on the total surface density when solving the \enquote{inverse} problem. It is thus not a surprise that the surface densities we derive in this work are inversely proportional to $x_\mathrm{CO}$, as it is clear for example from \autoref{eq:rho0_isothermal}. Unless otherwise specified, from now on we will use the standard value $x_\mathrm{CO}=10^{-4}$, as derived from the ISM. There is however a lot of evidence that CO is depleted in proto-planetary disks (see e.g. \citealt{MiotelloPPVII} for a review). We will discuss in section \ref{sec:co_depletion} whether this value is justified for proto-planetary disks.

\section{Calibrating the photo-dissociation column $N_{ph}$}
\label{sec:calibration}

As mentioned in the previous section, before we can solve the \enquote{inverse} problem and derive the disk surface density given the observed emitting height, there remains a free parameter to be estimated: the photo-dissociation column $N_{ph}$. Determining this parameter is the aim of this section. In order to accomplish this goal we compare our model, solving the \enquote{direct} problem, to the grid of thermochemical models presented in \citet{Teresa2024}.

\subsection{DALI thermo-chemical models}
While we refer to \citet{Teresa2024} for details, we briefly describe here the models we employ, which correspond to the fiducial parameters in \citet{Teresa2024}. Our motivation for only considering fiducial parameters is their conclusion that the parameters mostly affecting the emitting height are the disk mass and CO abundance, while the other parameters have a more limited effect. The models have been run using the DALI thermochemical code \citep{Bruderer2013}, which solves self-consistently for the (dust and gas) temperature and the chemical abundances of the various molecular species. We particularly stress that for this work we use the chemical network from \citet{Miotello2014}, which crucially accounts for CO isotope-selective effects (isotope-selective photodissociation and fractionation reactions) in addition to self-shielding and freeze-out. This is of fundamental importance for our work because, due to isotope-selective effects, the relative abundance of $^{12}$CO and $^{13}$CO can in principle differ from the overall isotopic ratio. This is a particular concern close to the photo-dissociation region, since photo-dissociation is an isotope-selective process.

The surface density of the disk is assumed to follow a self-similar solution \citep{LyndenBellPringle1974}:
\begin{equation}
    \Sigma (R) = \Sigma_c \left( \frac{R}{R_c}\right)^{-p} \exp \left[ -\left(\frac{R}{R_c}\right)^{2-p} \right],
\end{equation}
where $\Sigma_c$ is a normalization factor, $R_c$ a scale radius and $p$ a free parameter describing the steepness of the surface density. In this work we restrict to the case $p=1$ while we consider $R_c$=100 au and we consider four different values for $\Sigma_c$, setting it so that the total mass is $[0.0001, 0.001, 0.01, 0.1]\rm M_\odot$. We include both a population of small and large grains; the large grains dominate the mass, containing 95 per cent of the dust mass. The total dust-to-gas ratio is 0.01. The small grain population is assumed to be well coupled to the gas, while the large grain population is vertically settled, with a settling parameter (which should be interpreted as the ratio between the dust and gas scale-height, see \citealt{Bruderer2013}) $\chi=0.2$. As illuminating spectra, we consider both a T-Tauri star with 1 $L_\odot$ and a Herbig star with 17 $L_\odot$. The disk structure is computed solving for hydrostatic equilibrium; an iterative procedure \citep{Trapman2017,Teresa2024} is performed to ensure that the density and temperature structures are consistent with each other. The typical run time for the complete iteration process varies between 6-10 days for each full disk model, depending on the disk mass, underscoring why a simpler, but still effective, model is desirable to interpret observations. Finally, for each model, we ray trace the $^{12}$CO and $^{13}$CO emission and compute the height where the vertical optical depth at the line center is 2/3.


\subsection{Comparison between the emitting heights from DALI and from the \enquote{direct} problem}

\begin{figure*}
    \includegraphics[width=\textwidth]{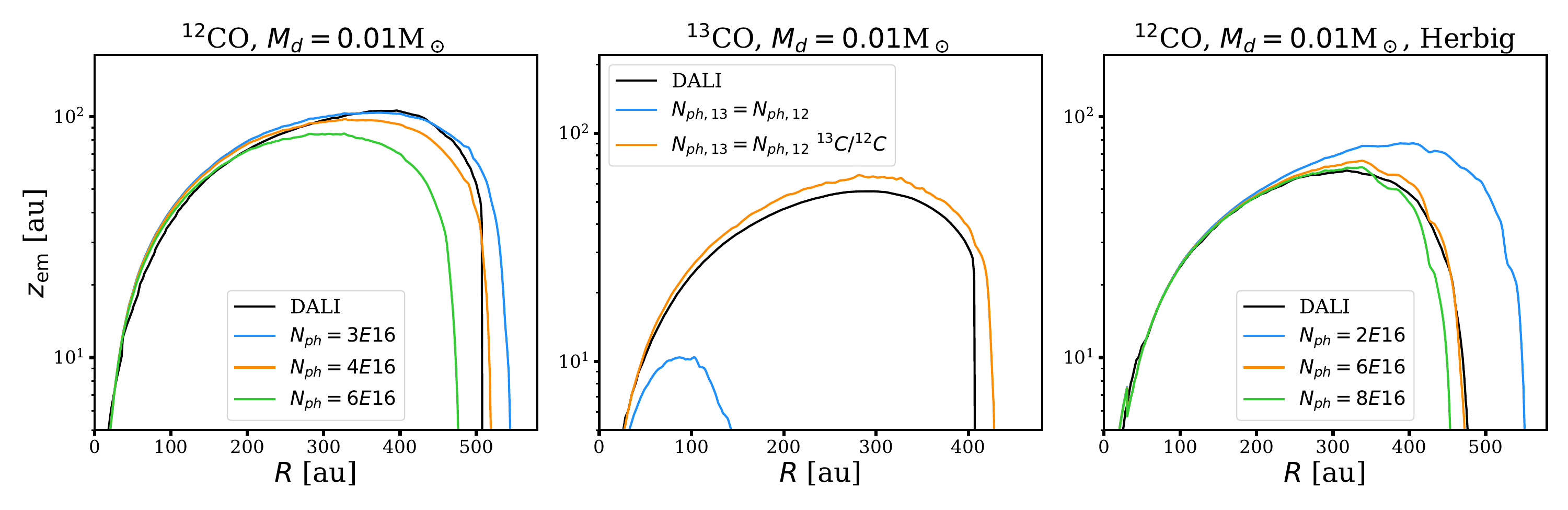}
    \caption{Left panel: Emitting height as a function of radius for $^{12}$CO, assuming a disk mass of 0.01 $M_\odot$. We show the values from the DALI thermo-chemical model, as well as from our semi-analytical model, for different values of the free parameter $N_\mathrm{ph}$. Central panel: Emitting height as a function of radius for $^{13}$CO, assuming a disk mass of 0.01 $M_\odot$ for different values of $N_{ph}$. We compare the DALI data (black line) with the emitting layer predicted assuming $N_{ph,13}=N_{ph,12}$ (blue line) and $N_{ph,13}=N_{ph,12}$ $   ^{13}\mathrm{C}/^{12}\mathrm{C}$ (orange line). Right panel: Emitting height as a function of radius for $^{12}$CO, assuming a disk mass of 0.01 $M_\odot$ orbiting a Herbig star. We show the values from the DALI thermo-chemical model, as well as from our semi-analytical model, for different values of the free parameter $N_\mathrm{ph}$.}
    \label{fig:comparison_dali}
\end{figure*}

We now compare the emitting heights between DALI and the solution of our direct problem. To make sure this is a fair comparison we read in the temperature computed by the DALI model and use it in our approach. We also use the DALI total mass density structure, which already accounts for hydrostatic equilibrium, rather than solving for it explicitly \footnote{We verified that using our solver for hydrostatic equilibrium we get a similar density profile to DALI. Some minor difference arises because of the DALI iterative procedure, which means that the density structure is computed solving hydrostatic equilibrium with the temperature computed at the previous step. This effect however is small and negligible for our purposes.}. Note this can correspond to a different CO volume density, since in our approach the CO abundance is parameterized, while in DALI it is computed self-consistently from the chemical network.

We show in the left panel of \autoref{fig:comparison_dali} an example with the results of the comparison for the case with $M_\mathrm{disk}=0.01 M_\odot$ for $^{12}$CO around a T Tauri star. The figure shows the emitting height as a function of radius for the DALI model and for our model for different values of the free parameter $N_{ph}$. The figure shows that in general our model matches well the results of the more accurate, but also significantly more computationally expensive, DALI model, and therefore serves also as a validation that our model, while simplified, correctly captures the most important aspects of the problem. However, the exact result depends on the assumed value of the self-shielding column. As expected, \textit{increasing} this value \textit{lowers} the emitting height, since the photo-dissociation layer becomes thicker. A simple inspection of the figure leads to a value of $N_{ph} \sim 3-4 \times 10^{16} \ \mathrm{cm}^{-2}$ as the one minimizing the difference between our model and DALI. To be more quantitative, we computed the mean squared deviation between our and DALI models and looked for the value of $N_{ph}$ that minimizes it. For this particular case we settle on a value of $N_{ph}=4 \times 10^{16} \ \mathrm{cm}^{-2}$. We repeated the analysis for the other values of the disk mass, finding in general similar values, although we do find that $N_{ph}=3 \times 10^{16} \ \mathrm{cm}^{-2}$ is a better fit for the case with $M_\mathrm{disk}=0.0001 M_\odot$. In a related problem, determining the radius beyond which CO is photo-dissociated, \citet{Trapman2023} also found that the critical gas column to shield against photo-dissociation gets smaller for lower disk masses. Given the relatively modest change in this case, we ignore this difference in what follows.

The case of $^{13}$CO deserves a separate discussion. Simple tests show that using the same value we used for $^{12}$CO leads to severely underestimate the emitting height. We show this case in the middle panel of \autoref{fig:comparison_dali} as the blue line; the model is the same as in the left panel, but we now consider the $^{13}$CO emitting layer. This assumption corresponds to consider only shielding by $^{13}$CO itself. We find instead that for $^{13}$CO rescaling $N_{ph}$ with the isotopic abundance, employing a value $N_{ph,13}=N_{ph,12}  \ ^{13}\mathrm{C}/^{12}\mathrm{C}$, leads to a much better agreement with the DALI models, as shown by the orange line in the middple panel of \autoref{fig:comparison_dali}. This implies that in practice isotope-selective photodissociation is not particularly effective for $^{13}$CO, a finding in line with previous dedicated studies \citep[e.g.][]{Visser2009,Miotello2014} which they attributed to isotope-exchange reactions.

{Since we have DALI models with a Herbig star as the central source, we repeated the test to verify whether the optimal value of $N_{ph}$ changes significantly compared to the T-Tauri case, as it could be expected from the fact that Herbig stars have a significantly more intense UV field, potentially leading to enhanced CO photo-dissociation. We show our results in the right panel of \autoref{fig:comparison_dali}. We find that the value of $N_{ph}$ minimizing the discrepancy between our model and DALI remains consistent with the T-Tauri scenario, with a slight increase to $6 \times 10^{16}$ cm$^{-2}$. We use this value for consistency, although we note that in practice the difference is small.}

We expect that other parameters that we did not explore here, such as for example $f_\mathbf{large}$, the fraction of large grains comprising the dust, could have an effect on $N_{ph}$ comparable the difference between T-Tauris and Herbigs, since they affect the amount of UV extinction by the dust. Although it is a slightly different problem, we note that also \citet{Trapman2023} found a weak (a factor $\sim 2$) dependence of the CO dissociation column on this parameter. A detailed exploration of how $N_{ph}$ depends on the microphysics is however outside the scope of this paper.

\section{Application to observations}
\label{sec:results}

In this section we describe how we apply the model we presented in the previous sections to interpret measurements of the emission height of proto-planetary disks. As part of the exoALMA paper series, in this paper we gather most of our sample from the exoALMA sample. In addition, we consider also the disks included in the MAPS \citep{ObergMAPS} programme since they also have high quality kinematical data available. For what concerns the exoALMA sample (see \citealt{exoALMA:Richard} for an overview), measurements of the emitting height are available using two different techniques, \discminer \citep{exoALMA:Andres} and \disksurf \citep{exoALMA:Maria}. The latter paper presents a comparison between the two measurements and a benchmark based on radiative transfer models. In general the two techniques yield similar results, but nevertheless the differences are notable. Based on the results of the comparison with radiative transfer models, which show that \discminer typically reconstructs the emitting height with higher accuracy at high inclinations due to less confusion with the backside, we opt to use the \discminer surfaces. This has the disadvantage though that we must work with parametrized surfaces. We show in appendix \ref{sec:ds_dm} how the results change when using \disksurf.

For what concerns exoALMA, we are not able to apply our model to the whole sample. We discarded in particular sources that show strong non-axisymmetric features (MWC 758, CQ Tau), since in our model we implicitly assume azimuthal symmetry, and low inclination ones (HD 135344B, HD 143006, J1604), for which the emitting layer cannot be reliably extracted. Finally, we exclude AA Tau for which the emitting layer extraction is problematic, most likely due to the high inclination. Therefore, the sample used in this work includes DM Tau, HD 34282, J1615, J1842, J1852, LkCa15, PDS66, SY Cha and V4046. Our requirements are very similar to the analysis of rotation curves presented in \citet{exoALMA:Cristiano} and our samples mostly overlap, with the exception of AA Tau for which they still present their results (though they caveat they may be inaccurate). For what concerns the MAPS sample (AS 209, MWC 480, HD163296, IM Lup and GM Aur), for consistency we use the measurements of the emitting heights described in \citet{IzquierdoMAPS} that also used \discminer, rather than the surfaces originally reported by the MAPS collaboration \citep{Law2021} that were derived with \disksurf. Combining the exoALMA and MAPS sample, we have in total 14 disks.

As previously mentioned, we need to rely on an independent measurement of the temperature to apply our method. For the MAPS sample, we adopt the profiles reported by \citet{Law2021}, while for the exoALMA sample the profiles reported by \citet{exoALMA:Maria} - note that the two works assumed a different functional form for the temperature profile. For what concerns $^{13}$CO, \citet{exoALMA:Maria} reports that in the outer parts of the disk the brightness temperature can be lower than 20 K. Since at face value this points to the emission no longer being optically thick, we rescale in this case the critical optical depth traced by the emitting layer (see \autoref{eq:z_em_implicit}) by the ratio between the brightness temperature and 20 K. We find however this is only a small correction (a 10\% factor on average on the total mass). In terms of radial range, we only consider radii larger than 0.3" since the inner part of the disk is affected by beam smearing. For the MAPS sample, \citet{Law2021} reported radial ranges where their fits are valid and we therefore apply the further constraint that the radius must be contained inside the reported range. For the exoALMA sample, we use the parameter $R_\mathrm{out}$ reported by the \discminer fit to set the outer radius. Since $^{13}$CO emission is typically less extended than $^{12}$CO, we are in general able to derive a surface density on a wider radial extent for $^{12}$CO than we are for $^{13}$CO. We find that in most cases (see also section \ref{sec:compare_12_13}) there is negligible mass in the region where we get constraints only from $^{12}$CO.

A final point to discuss when dealing with observations is which errors to report for our estimates. Since our method is computationally cheap, we can afford to bootstrap the errors on parameters inputs. For the temperatures, we simply use the uncertainties reported by \citet{exoALMA:Maria} and \citet{Law2021} for their temperature fits. For the emission surfaces, however, we cannot directly use the errors reported by \discminer; they are likely to be severely underestimated as a result of the strong systematic choice of a parameterized emission surface and using them directly we would obtain a severely underestimated error. Instead, we choose to use a representative uncertainty based on benchmark performed on radiative transfer models (see appendix of \citealt{exoALMA:Maria}). We find that 25 per cent is a typical uncertainty and we therefore use this value in the rest of the paper. Because of this choice, we remark that the uncertainties we show should not be considered as true statistical uncertainties but should instead be considered as representative. Operationally, we assume the error is normally distributed and we bootstrap 1000 times to quantify uncertainties. Since the results can vary by orders of magnitude, the values we report in this paper are computed from the average and standard deviation in log space of the resulting 1000 samples.


The two measurements of $^{12}$CO and $^{13}$CO emitting heights correspond to two different measurements of the disk surface density when using the stratified model - we will discuss their (dis)agreement in section \ref{sec:compare_12_13}. For what concerns the isothermal model, as discussed in section \ref{sec:model}, there are in principle multiple options depending on which temperature we use. Combined with the stratified case, this leads potentially to a large number of surface density measurements and it needs to be discussed which of these are the most reliable. To simplify the illustration of our results, we now focus mostly on the stratified model, which we consider our best estimate. We discuss in section \ref{sec:comparei} the results of the isothermal model when we use the line brightness temperature. We anticipate the results in this case are relatively comparable to the stratified case, while we find instead that solving the system of two equations to derive $H$ cannot be always reliably applied. For this reason we discuss this case in appendix \ref{app:system_two_equations}.


\subsection{A worked example: LkCa 15}
\label{sec:im_lup}

\begin{figure}
    \includegraphics[width=\columnwidth]{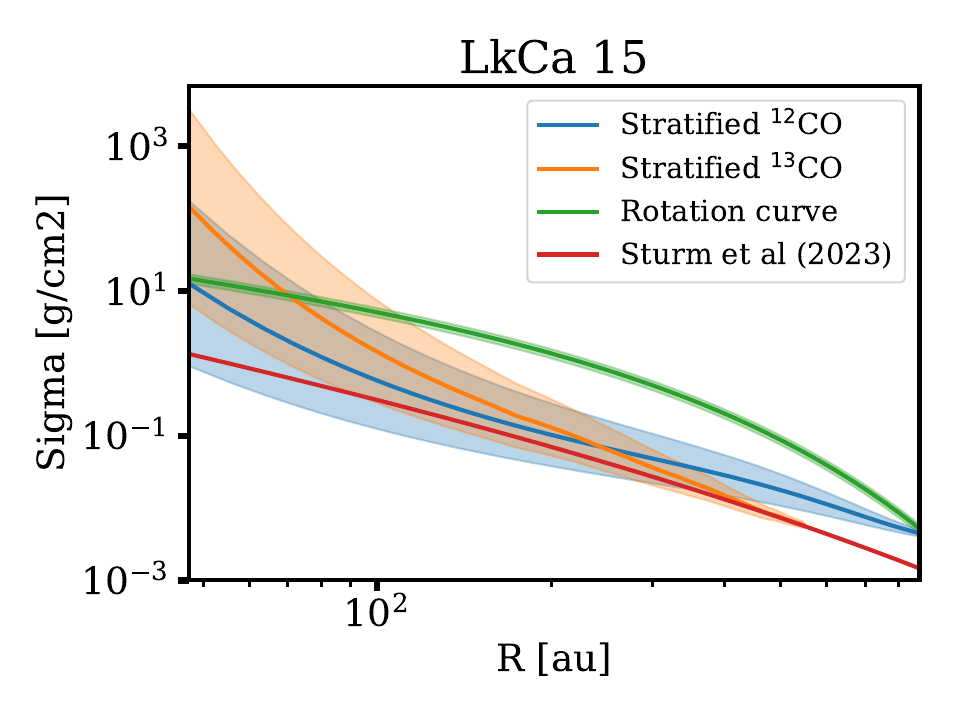}
    \caption{Surface density as a function of radius derived with the stratified model for LkCa 15. We show the results for both $^{12}$CO and $^{13}$CO; shaded regions show the uncertainties (see text for a discussion of the radial dependence of the uncertainty). For comparison we also show the surface density implied by the fit to the rotation curve presented in this paper series in \citet{exoALMA:Cristiano} and a constraint from optically thin C$^{18}$O and C$^{17}$O lines presented in \citet{Sturm2023}.}
    \label{fig:results_lkca15}
\end{figure}

Given the large number of surface density measurements when applying the various combinations in our method to the whole sample, we now choose to focus on a specific example that we illustrate in detail. As representative example we choose LkCa 15, a large and bright disk that thus constitutes one of the best case scenarios for applying our method; in addition it also has independent constraints from optically thin lines (C$^{18}$O and C$^{17}$O), presented in \citet{Sturm2023}, that we can compare to. \autoref{fig:results_lkca15} shows the fitted surface density. We also show the surface density profile resulting from the disk mass reported in the accompanying paper by \citet{exoALMA:Cristiano}, obtained by fitting the disk rotation curve. We remark that, while the method we employ here makes no assumption about the shape of the surface density profile, \citet{exoALMA:Cristiano} and \citet{Sturm2023} assumed a self-similar profile.

We first note that the surface density profiles we derive with our method appear to be in a reasonable range - their values are bracketed in most of the radial range by the other two constraints. The implied surface density profile decreases with radius, as one may naively expect. As already mentioned our method can be essentially considered as an extrapolation of the density from the emitting height down to the midplane, and as such it yields large uncertainties; for reference, at a radius of 100 au the uncertainty is 0.7 dex. The advantage of our method is that it can be applied to optically thick emission, but this comes at the cost of large uncertainties. We also note that the uncertainty decreases with radius. This is driven by the decreasing trend we described of the surface density with radius; a higher surface density implies that the emitting height is in a region where the density gradient with height is steep, causing larger uncertainties in the extrapolation. Finally, we find that the uncertainty is entirely dominated by the uncertainty in the emitting height determination; including only the error on the temperature produces an uncertainty that is not visible on the scale of the plot. This is because the errors reported by \citet{exoALMA:Maria} are in general small (e.g., fractions of a K for the normalizations).

Comparing $^{12}$CO and $^{13}$CO, we note that the values are close to each other and are compatible within the uncertainties for most of the radial range. More quantitatively, the total disk mass derived from $^{12}$CO is $0.02 \ M_\odot \pm 0.7 \mathrm{dex}$ (i.e., $0.02^{+0.08}_{-}$) and $0.06 \ M_\odot  \pm 1 \mathrm{dex}$ for $^{13}$CO. Given the uncertainties, the two estimates are in good agreement. The value reported in \citet{Sturm2023} is slightly lower at $0.01 M_\odot$. From the figure it could seem that this estimate should be even lower, but this is because the mass in this case is integrated over the whole domain; integrating over only the radial extent where we get a surface density measurement from $^{12}$CO yields a factor 2 lower mass.

A simple fit with a power-law surface density for $^{12}$CO returns a value of -2.3 for the exponent. This value is steep but it should be interpreted with caution since a comparable fit to the optically thin constraint yields a similar value of -2.4; this is due to the exponential drop off of the disk surface density in the outer regions. 

Our estimates are lower than the dynamical mass ($\sim 0.1 \ M_\odot$), and this is significant for $^{12}$CO. Bearing in mind that in order to derive our estimates we had to assume a functional form of the temperature structure, which in principle could affect our results, we interpret the difference as due to CO depletion. Since our constraints are derived assuming the standard ISM CO abundance of $10^{-4}$, the implication is that in this source CO must have a factor $\sim$ 10 lower abundance. We will show this conclusion is not an isolated case but holds for the whole sample (and in fact LkCa 15 is the source in the sample with the smallest amount of depletion).

\subsection{Surface density profiles for the whole sample}
\label{sec:surface_density_everything}

\begin{figure*}
    \centering
    \includegraphics[width=.32\textwidth]{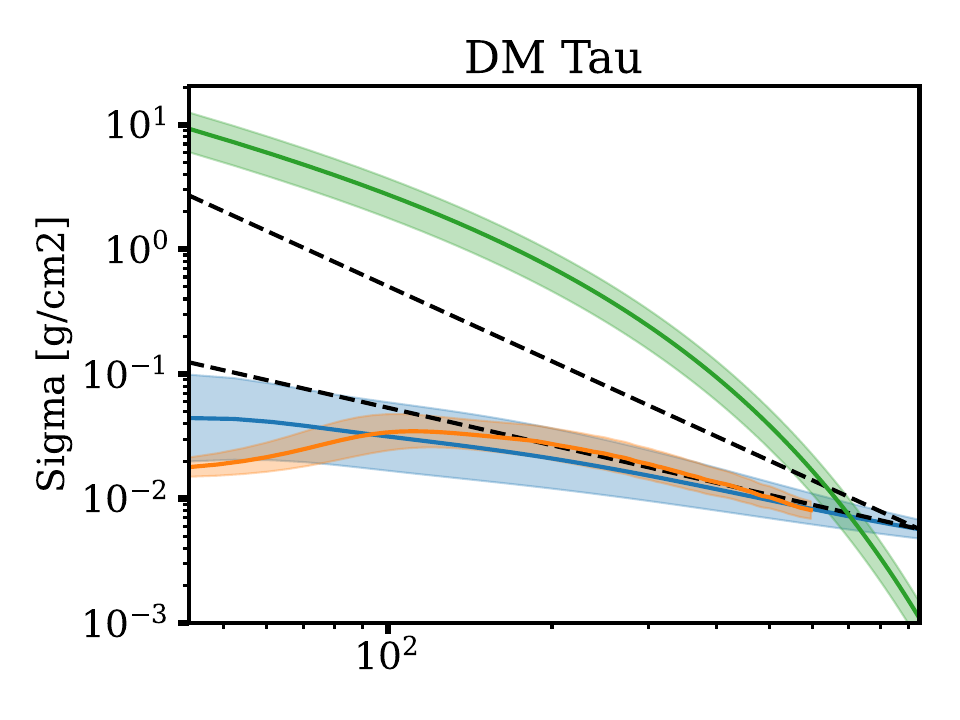}
    \includegraphics[width=.32\textwidth]{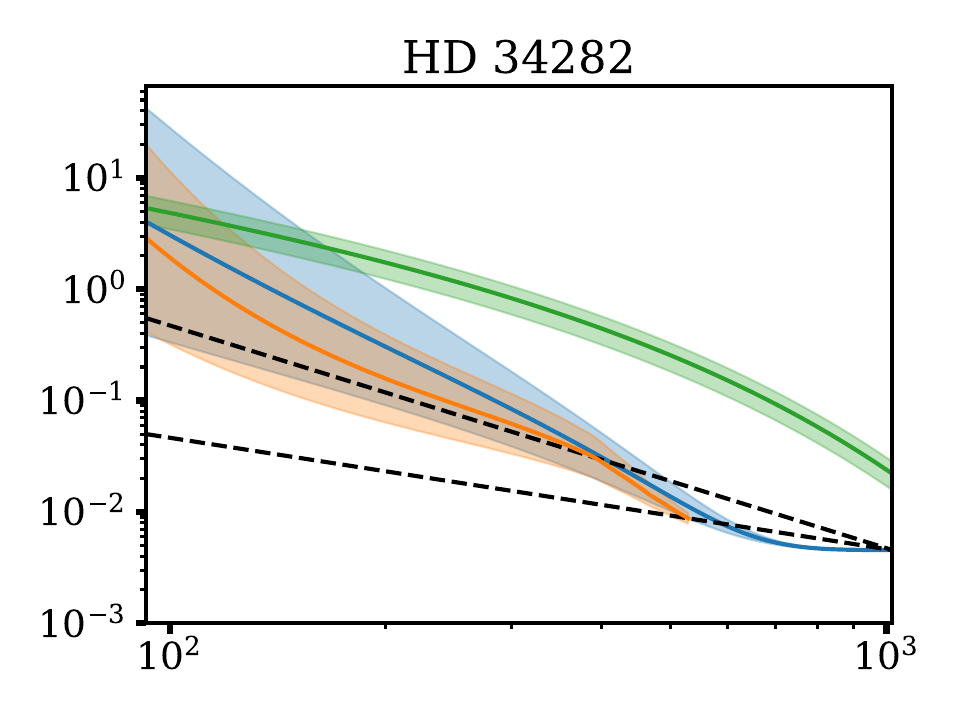}
    \includegraphics[width=.32\textwidth]{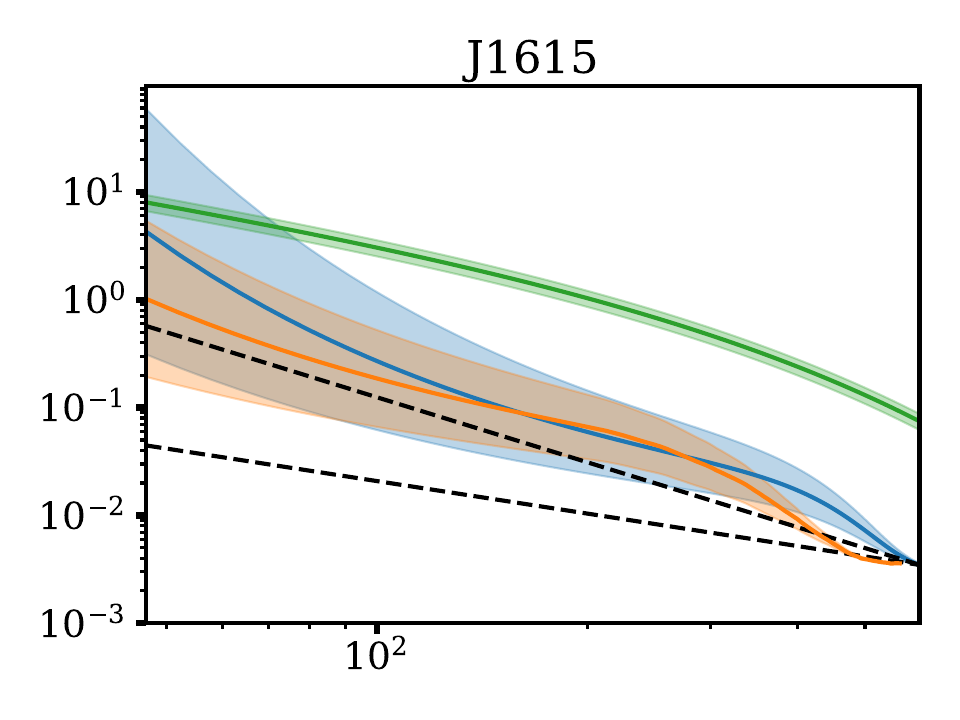}
    
    \includegraphics[width=.32\textwidth]{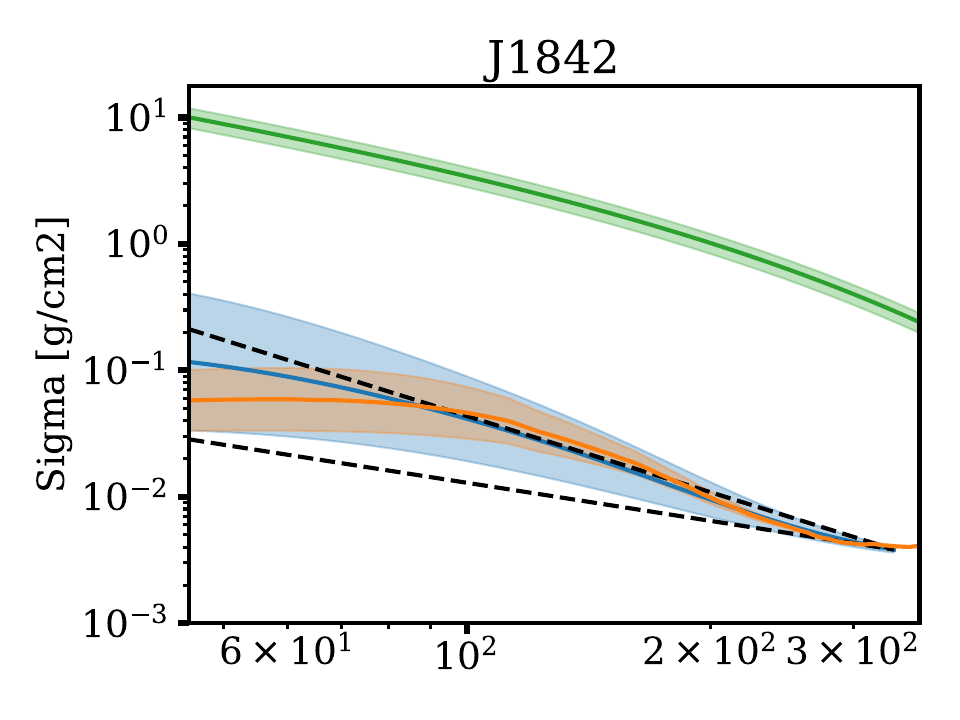}
    \includegraphics[width=.32\textwidth]{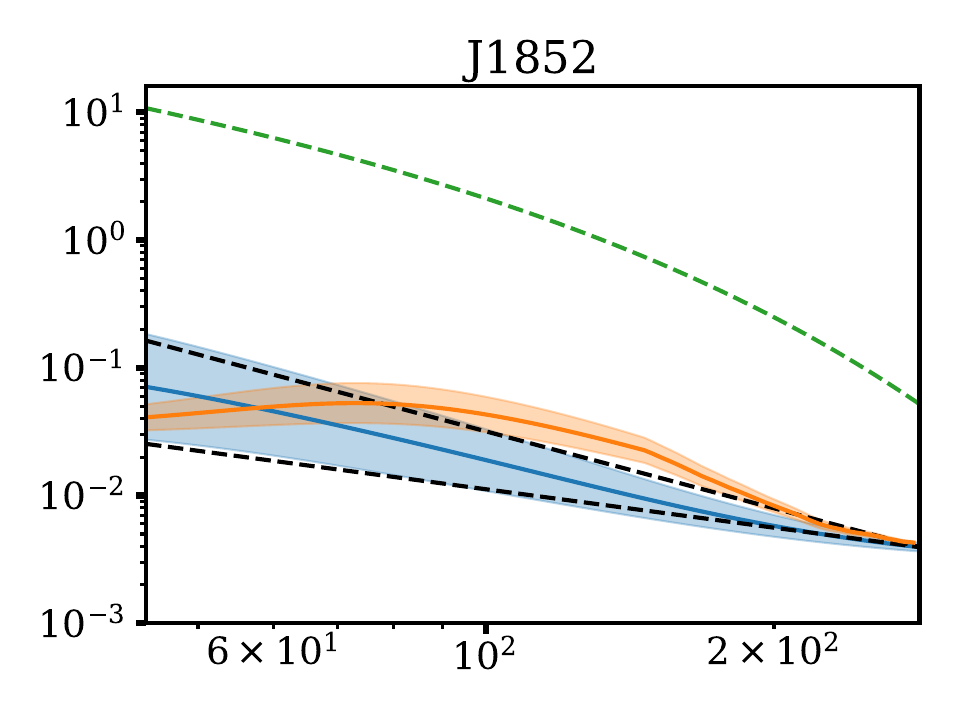}
    \includegraphics[width=.32\textwidth]{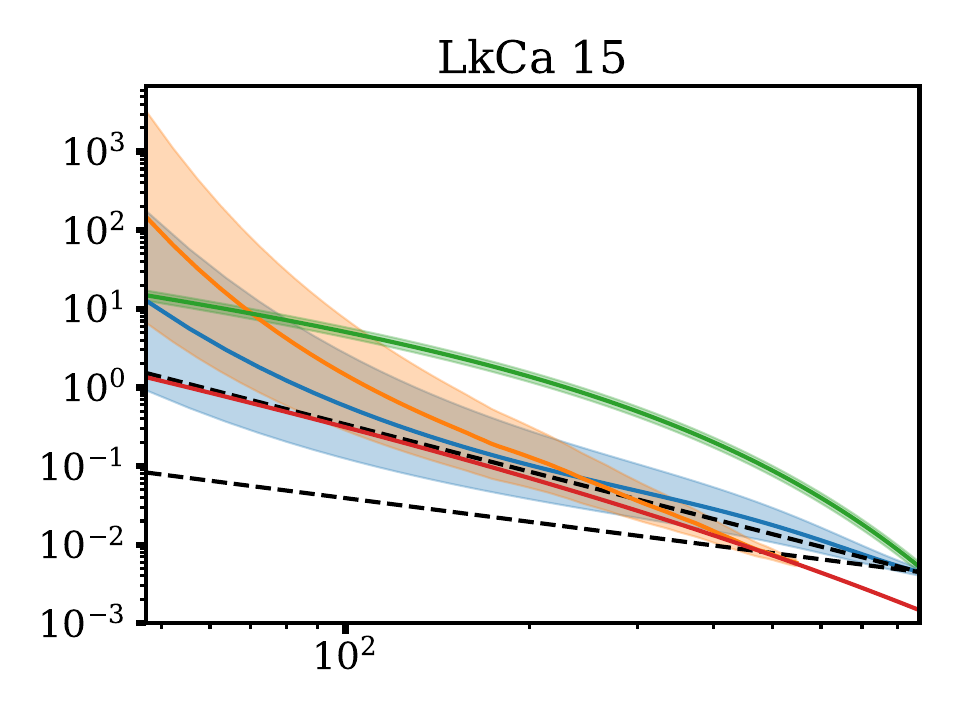}
    
    \includegraphics[width=.32\textwidth]{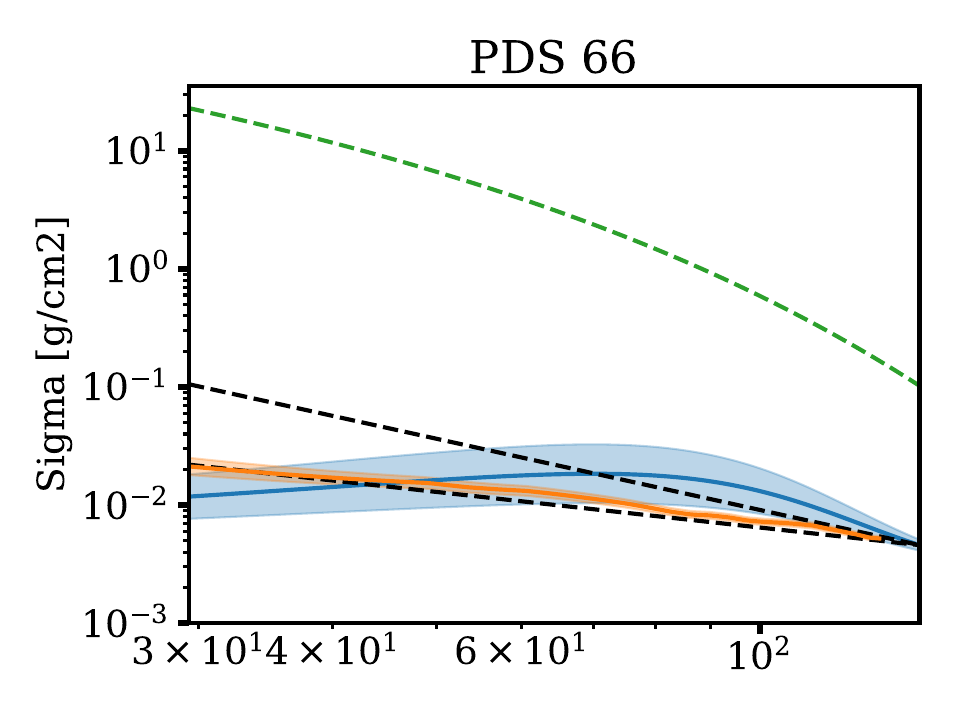}
    \includegraphics[width=.32\textwidth]{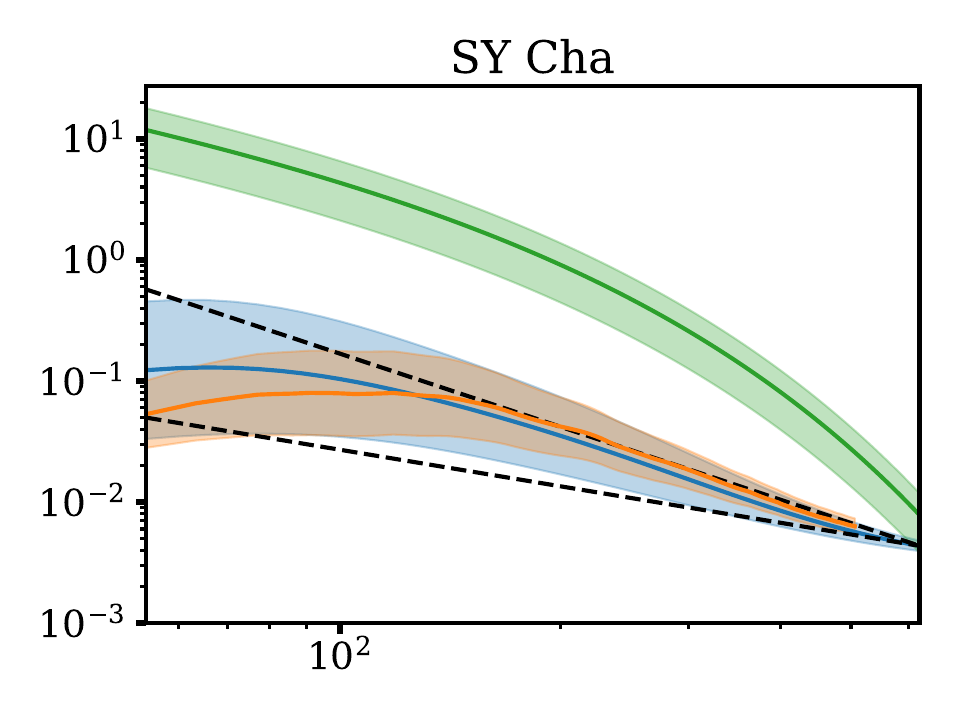}
    \includegraphics[width=.32\textwidth]{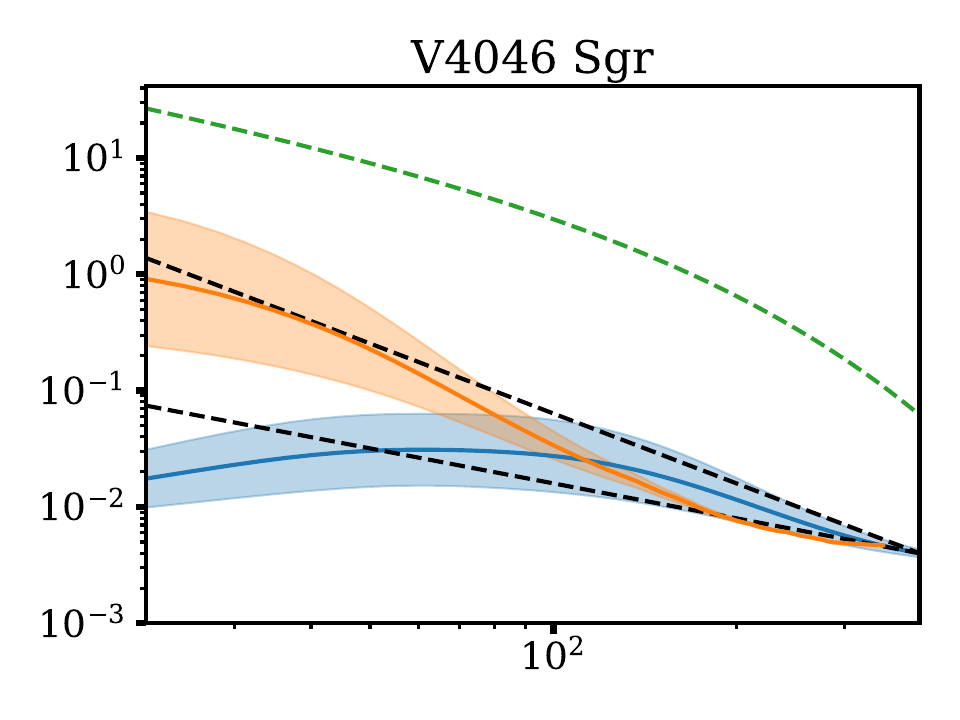}
    
    \includegraphics[width=.32\textwidth]{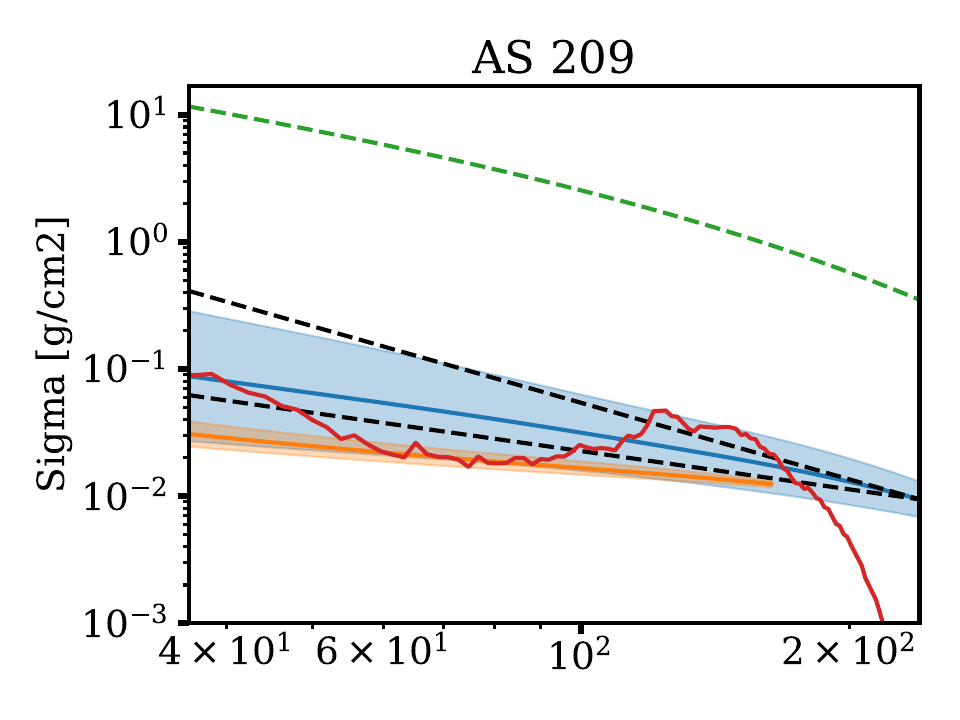}
    \includegraphics[width=.32\textwidth]{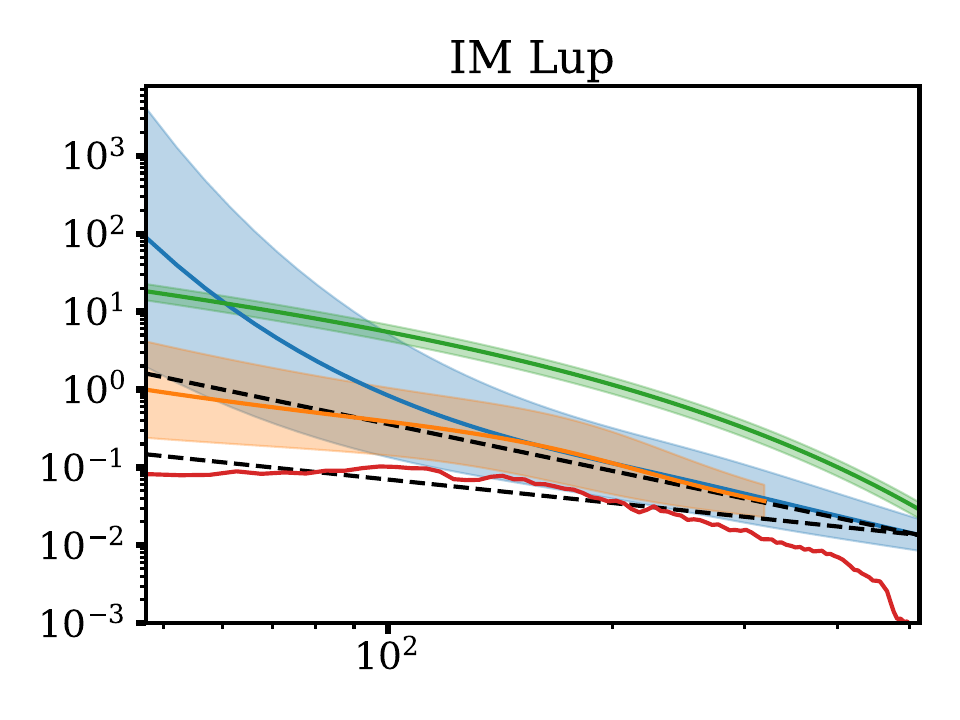}
    \includegraphics[width=.32\textwidth]{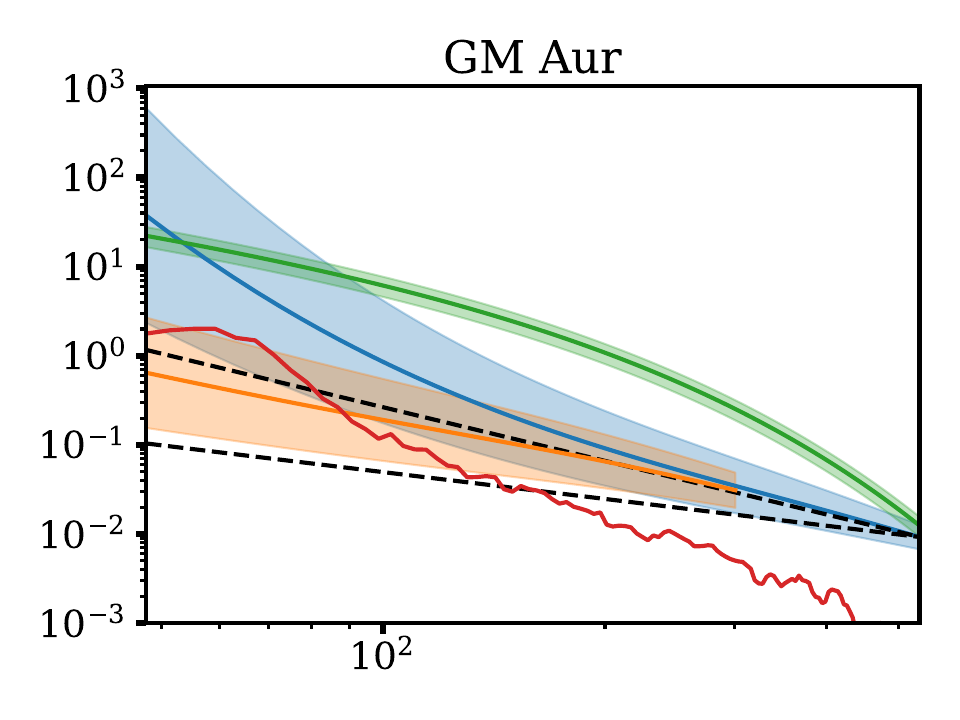}

    \includegraphics[width=.32\textwidth]{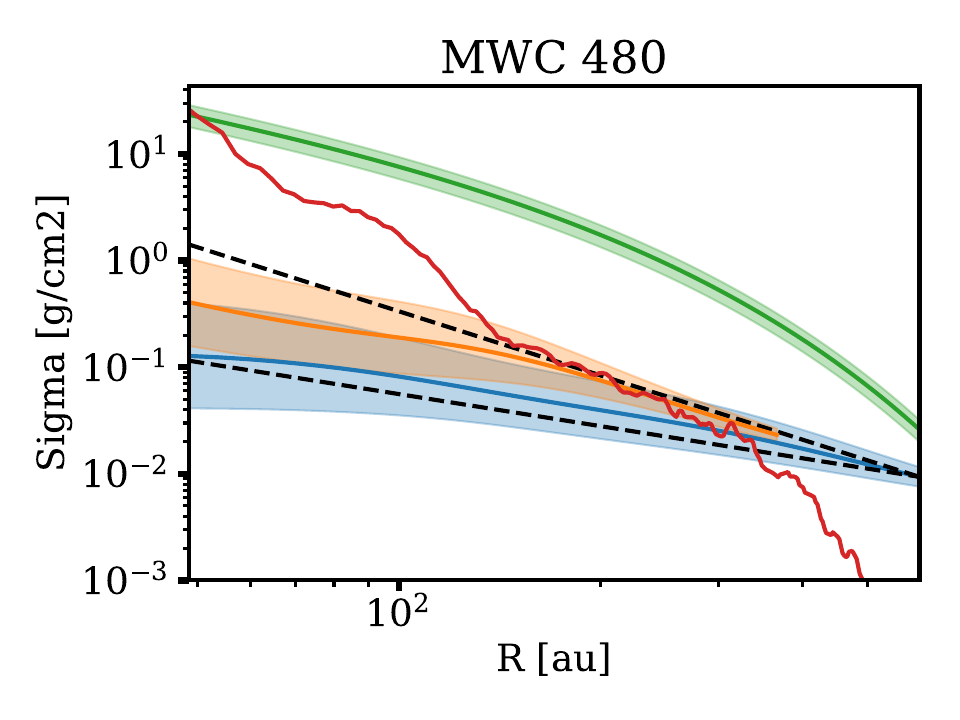}
    \includegraphics[width=.32\textwidth]{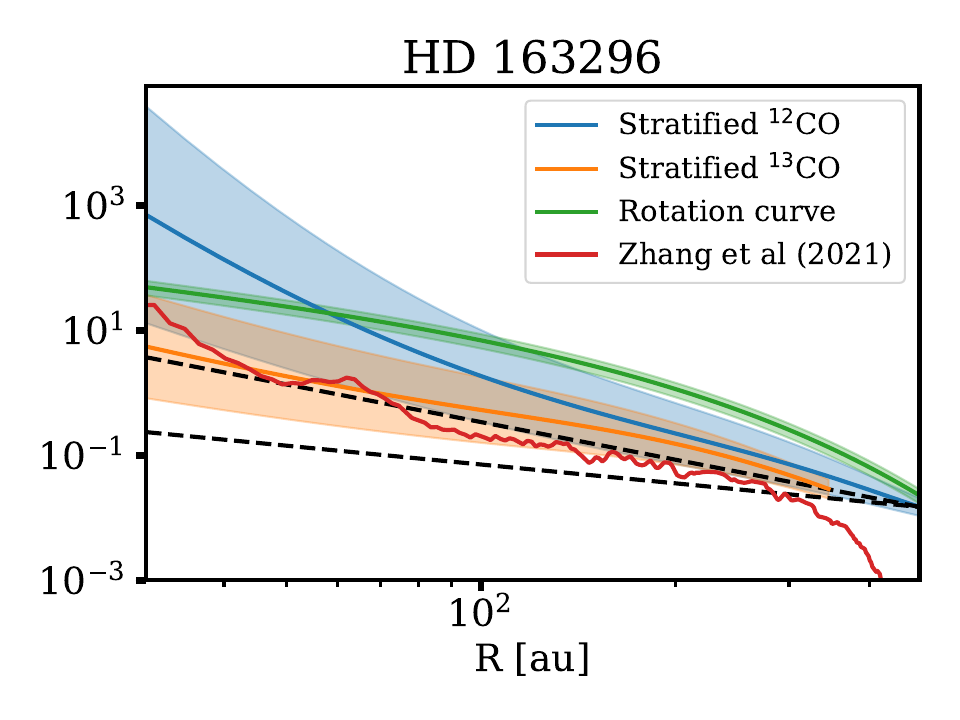}

    \caption{Like \autoref{fig:results_lkca15}, but for all the disks in our sample. We plot the constraint from the rotation curve \citep{Martire2024,exoALMA:Cristiano} as a dashed line when they only constitute upper limits. For the MAPS sample, the red line shows the surface density constraints from (partially) optically thin C$^{18}$O \citep{Zhang2021}. Black dashed lines are visual aids showing $r^{-1}$ and $r^{-2}$ profiles.}
    \label{fig:gallery_sigmas}
\end{figure*}

After describing LkCa 15 in detail, we now show the results for the whole sample in \autoref{fig:gallery_sigmas}. As for LkCa 15, we find that uncertainties are always dominated by the error we attribute to the emitting height determination, rather than the temperature uncertainty. While we will not describe each disk individually, we note that many of the trends we described before still hold. To start with, the surface densities we derive decrease with radius\footnote{Although some of our sources are transition disks and potentially have inner gas cavities, the cavity radius is inside the two central beams.}. The two estimates from $^{12}$CO and $^{13}$CO generally agree within the uncertainties, but they are always lower than the estimate from the rotation curve (reported in \citealt{exoALMA:Cristiano} for exoALMA and in \citealt{Martire2024} for MAPS). In the interpretation we discussed above, this points to CO depletion being a widespread phenomenon. We will quantify the level of depletion in the next section.  As for LkCa 15, we remark that the estimates derived from the rotation curve assume a self-similar profile with a fixed slope (but see \citealt{AndrewsRotationCurve} for MWC 480, where the slope is let free to vary). We also added as visual aids black dashed lines showing $r^{-1}$ and $r^{-2}$ profiles in order to gauge the steepness of the surface density slope. Most of the surface densities we derive follow more closely the steeper profile (see also the discussion in \citealt{exoALMA:Maria}).

Finally, for the MAPS sample, we also included as the red line the surface density constraints from (partially) optically thin C$^{18}$O \citep{Zhang2021}. We take their result on the CO column density and rescale it to a total density with the same assumption we employ to derive our constraints, i.e. a constant CO abundance of $10^{-4}$. We note this is different from the surface density reported in \citet{Zhang2021}, who also employed information from the SED to derive the total disk mass and in this way derive a spatially resolved CO depletion profile. We choose to compare our results in this way since the raw output of our method is a CO column, and comparing directly the surface density reported by \citet{Zhang2021} would not be a fair comparison.

\section{Discussion}
\label{sec:discussion}

\subsection{Comparison with the isothermal model}
\label{sec:comparei}

\begin{figure}
    \includegraphics[width=\columnwidth]{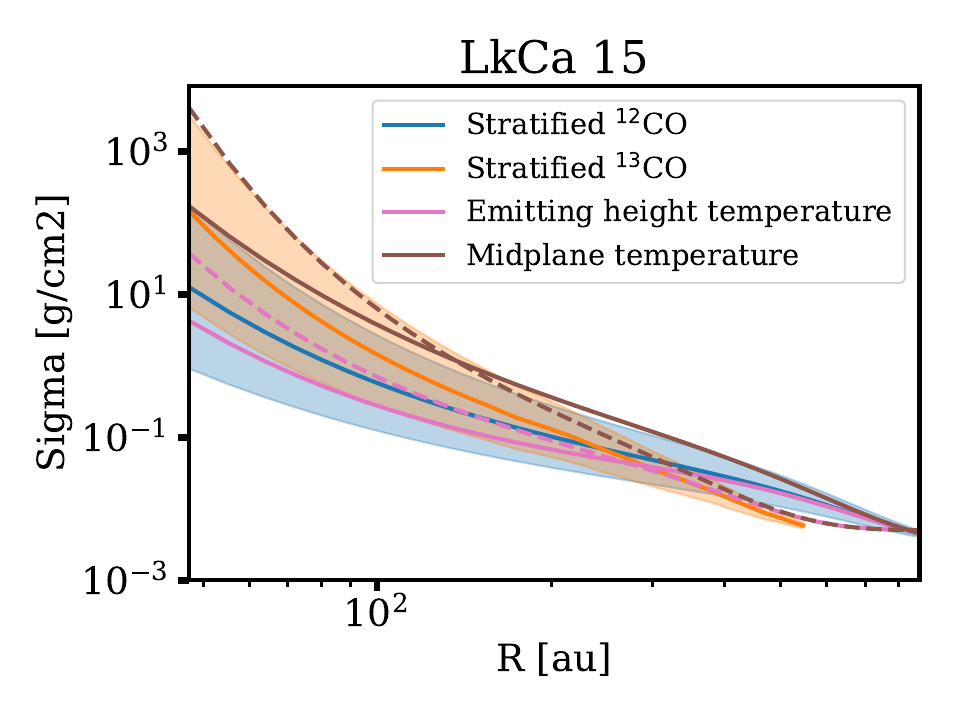}
    \caption{Like \autoref{fig:results_lkca15}, but we now show the results of the stratified model in comparison with the results of the isothermal model. We consider two cases, described in detail in the main text, in which we set the temperature based either on the brightness temperature of the line or on the fitted disk temperature profile in the midplane. For the isothermal model, solid lines are for $^{12}$CO while dashed lines are for $^{13}$CO.}
    \label{fig:comparei_lkca15}
\end{figure}

As mentioned, we consider the stratified model our best estimate since it uses the full information on how the temperature profile depends on the vertical coordinate. However, this approach requires having data for at least two emission lines. It is conceivable that in some circumstances high quality observations could be available instead for a single emission line, for example because one of the lines (in this case likely the most abundant, $^{12}$CO) is cloud contaminated, or because a line may be particularly faint (in this case likely the least abundant, $^{13}$CO), or simply because of requirements in the spectral setup that did not permit to include two. It is interesting in this case to quantify how well the surface density can be reconstructed. 

We show the comparison in \autoref{fig:comparei_lkca15} for the reference case of LkCa 15. For the isothermal model, solid lines are for $^{12}$CO while dashed lines are for $^{13}$CO. As discussed in section \ref{sec:inverse_isothermal}, we show estimates from the isothermal model using the brightness temperature at the emission surface to set the scale-height. In addition, for reference we include an additional estimate where we set the temperature using the temperature fit of \citet{exoALMA:Maria} in the midplane. This estimate cannot be computed when observations of a single line are available, and it is therefore of limited practical use. However, it is useful for illustrative reasons. First of all, we note that using the emitting height temperature gives an estimate that compares reasonably well with the more complex stratified case - for example, for $^{12}$CO the difference is roughly a factor 2. When using instead the midplane temperature, we see that it produces a much higher value than the stratified case (note the scale of the y axis) - to be more quantitative, integrating the surface density for $^{13}$CO leads to a problematic disk mass of $\sim 1 \ M_\odot$ when using the midplane temperature, compared to $0.07 \ M_\odot$ for the stratified case, and in a similar way for $^{12}$CO the difference is a factor 6.  We discuss more in detail the reasons why the midplane temperature tends to lead to an overestimate of the surface density, despite naively one may expect it to work better since most of the mass mass is at the midplane, in appendix \ref{app:extrapolation}.

\begin{figure}
    \includegraphics[width=\columnwidth]{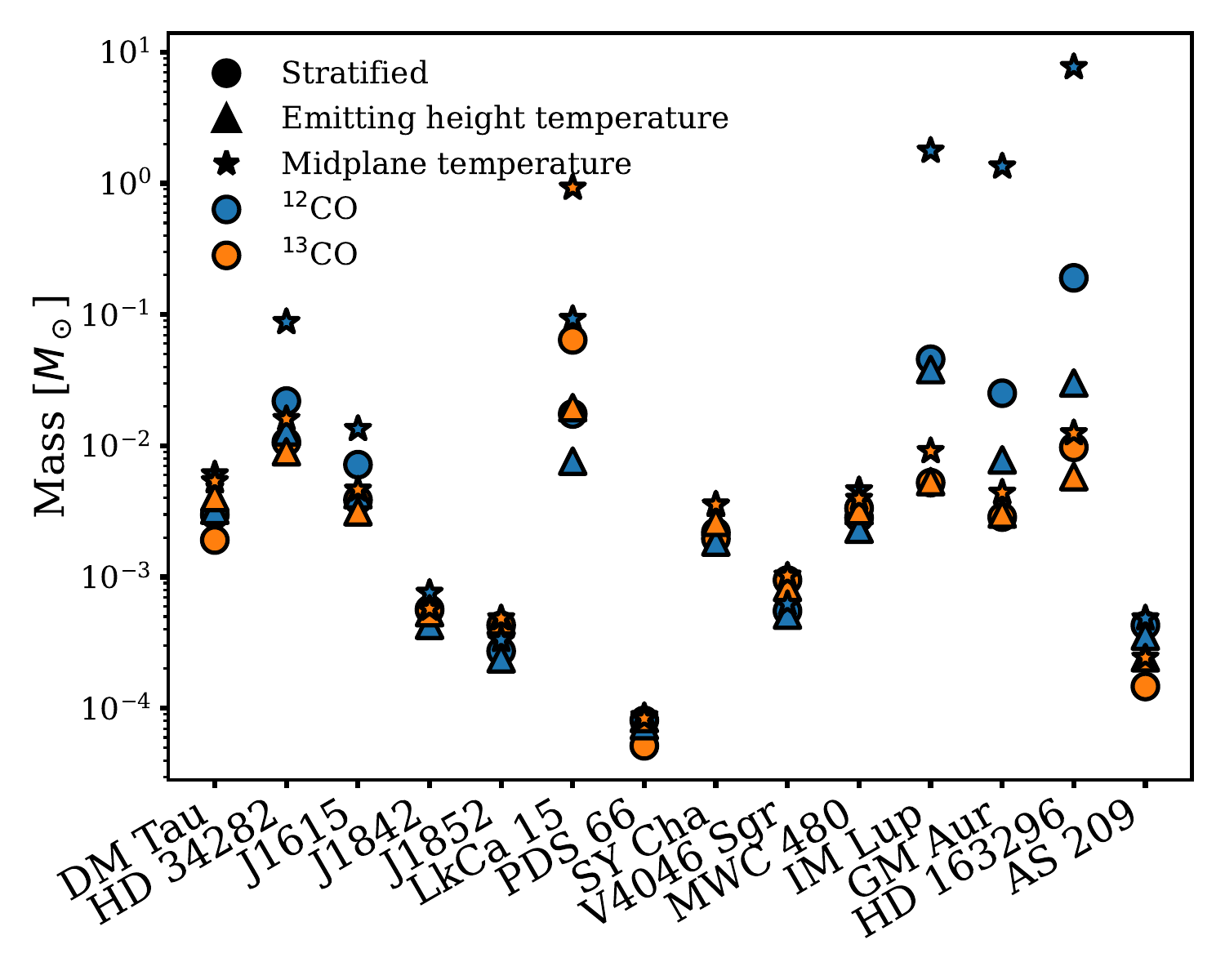}
    \caption{Masses derived from the isothermal model, assuming either the temperature in the midplane or the temperature at the emitting height, in comparison with the stratified case for the whole sample.}
    \label{fig:comparei_all}
\end{figure}

Moving to the whole sample, we show in figure \ref{fig:comparei_all} the results for all the disk masses. We do not show error bars on this plot since the various datapoints come from the same measurements and the goal is to compare the different techniques with each other. We can see that the trends discussed for LkCa 15 remain also for the whole population: the mass estimated using the midplane temperature always overestimates the temperature of the stratified model, sometime severely so (the average overestimate is a factor $\sim$6, but the plot shows in some cases it can be larger than one order of magnitude). On the other hand, using the emitting height temperature returns a value that most of the time is close to the stratified case, and in general underestimated. To quantify the discrepancy, we compute that the average ratio between the two masses is 0.8 (the median has a similar value) with a spread of $\sim$0.2 dex, which we can consider as an empirical uncertainty of the approximation. We conclude that, when only one line is available, using the isothermal model and using the brightness temperature at the emitting surface to set the temperature is a satisfactory approximation.

\subsection{Difference between $^{12}$CO and $^{13}$CO estimates}
\label{sec:compare_12_13}

\begin{figure}
    \centering
    \includegraphics[width=\columnwidth]{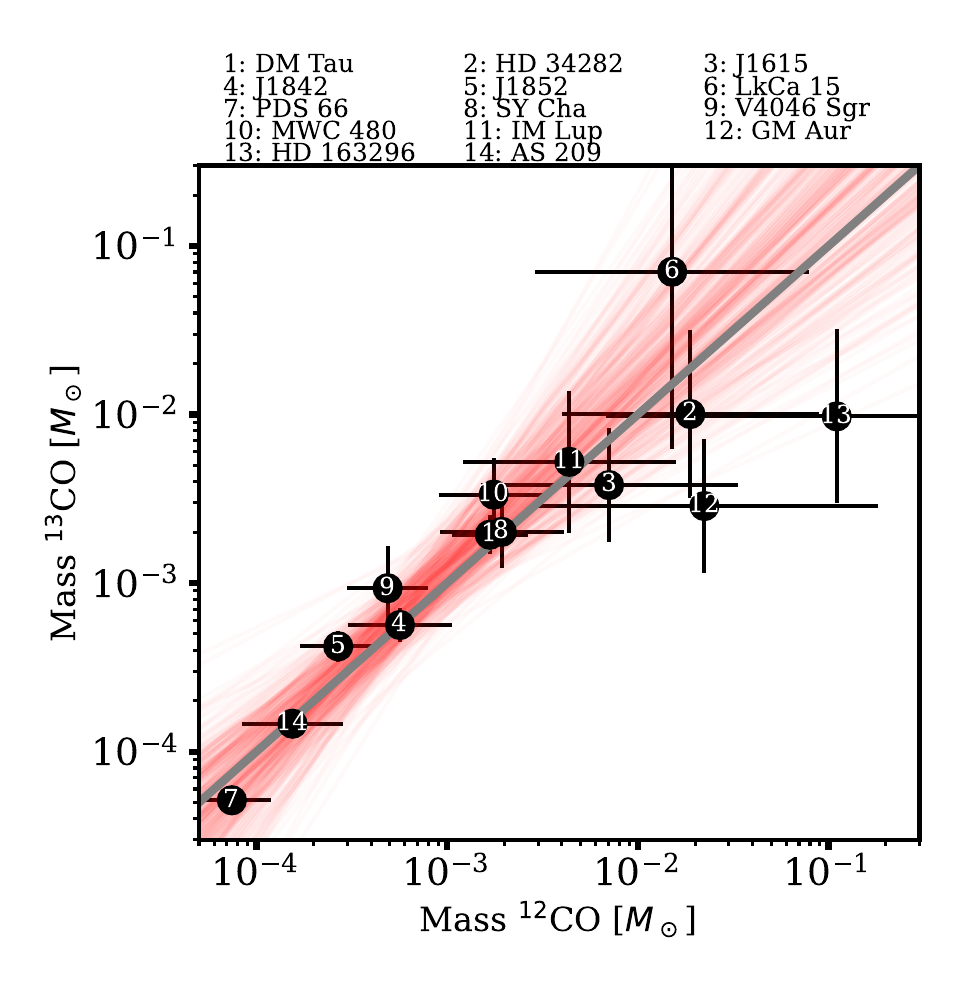}
    \caption{Comparison between the mass inferred using the $^{12}$CO emitting height and the mass inferred using the $^{13}$CO emitting height. The grey solid line marks where the two masses are the same. The red lines are the results of the power-law fit described in the main text.}
    \label{fig:mass_12_vs_mass_13}
\end{figure}

To make comparisons across the whole sample, we now refer only to the disk mass. In this subsection we investigate whether there is a systematic difference between the mass derived from $^{12}$CO and $^{13}$CO. We show the result of the comparison in \autoref{fig:mass_12_vs_mass_13}, in which we plot the mass derived from $^{13}$CO as a function of the mass derived from $^{12}$CO. For consistency, the $^{12}$CO mass reported in this plot has been computed only over radial range over which the $^{13}$CO estimate is defined, although this is a minor difference (a factor 2 on average, with the notable exception of IM Lup for which the difference is significant). The grey solid line represents equal values. The first thing to note is that there are a few outliers which are significantly far from the line where the two masses are the same. Two of these are HD163296 and GM Aur, for which the $^{12}$CO mass is higher than the $^{13}$CO. An inspection of the surface density reveals that the high $^{12}$CO mass is dominated by the inner regions of the disk where the $^{12}$CO surface density goes to very high values. LkCa 15 is instead in the opposite situation where the $^{13}$CO mass is significantly higher than the $^{12}$CO one. It is likely that in these disks there is an issue in the inner disk with one emission surface; however, the high surface density also implies large error bars, and these disks remain compatible with lying on the equality line. For this reason we decide to keep them in the sample.

For what concerns the other disks, there is overall good agreement between the two mass estimates, although it appears that on average the $^{13}$CO mass is slightly higher than the $^{12}$CO one. To quantify this, we fitted the values using a power-law relation with the method\footnote{An open source implementation is available at the following link: \url{https://github.com/jmeyers314/linmix}.} developed by \citet{Kelly2007}, which allows us to include errors on both axis. We show the results of the fit with the red lines. The fit confirms that there is a good correlation between the two values, with a slope of $1.1 \pm 0.2$, and also that on average the $^{13}$CO mass is higher than $^{12}$CO, with a value of the normalization $2.5 \pm 0.8$ dex. We note however that this value has a large error, meaning that the two mass estimates remain compatible within the uncertainties. We remark that a discrepancy could point to a isotopic ratio $^{12}$C/$^{13}$C different from the value of $\sim$77 we have assumed here. Indeed recently \citet{Yoshida2022} found evidence in TW Hya, where the almost face-on geometry allows to measure the column of material in the optically thin line wings, of a reduced $^{12}$C/$^{13}$C isotopic ratio when comparing $^{13}$CO to $^{12}$CO (see also \citealt{Bergin2024} for a discussion of C fractionation in this source). They measured a value of 21 at radii between 70 and 110 au. The value of the normalization we derive would correspond to a isotopic ratio of $\sim$30, but, considering the uncertainties in our estimates, we are unable to confirm whether we have evidence of the same behavior in our sample. It is worth remembering that our uncertainties are not true statistical uncertainties and could be over- or under-estimated, but since we regard them as typical we will not speculate further on the isotopic ratio. Further work is needed to either reduce the error bars or increase the sample to come to firmer conclusions.


Overall, the fact that the masses derived from the two isotopologues are in such good agreement is remarkable considering that our method relies on knowing the disk temperature structure and that we are estimating the mass from optically thick emission, although as we highlighted this comes at the cost of large error bars. We also note that the measurements of the two emitting heights are separate, so there is no guarantee \textit{a priori} that the two values be similar. 

\subsection{Comparison with other mass estimates and CO depletion}
\label{sec:co_depletion}

\begin{figure*}
    \centering
    \includegraphics[width=\textwidth]{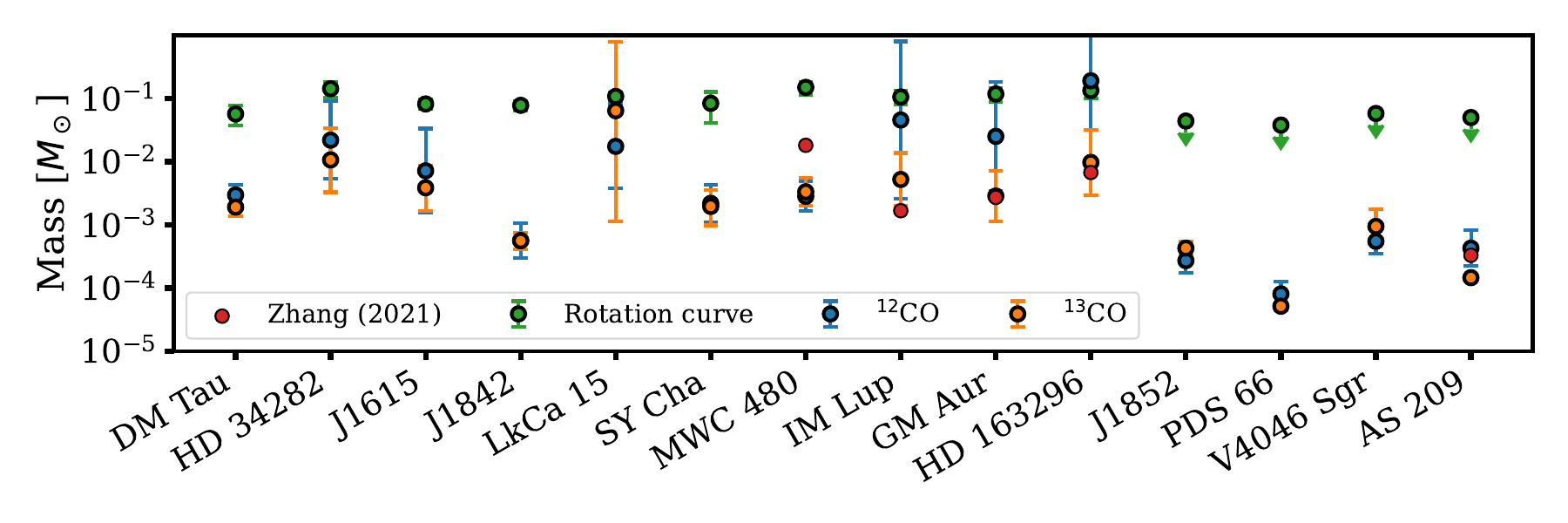}
    \caption{Comparison between the masses measured with various methods, namely the dynamical masses derived from the rotation curve \citep{Martire2024,exoALMA:Cristiano}, the emitting heights (this work) and (moderately) optically thin C$^{18}$O \citep{Zhang2021} for the disks in our sample.}
    \label{fig:comparison_masses}
\end{figure*}

\begin{table*}
    \centering
    \caption{Masses derived from $^{12}$CO and $^{13}$CO in this work, with their uncertainties.}
    \label{tab:masses}    
    \input{table_masses}
\end{table*}

We now compare the masses resulting from our method, which we list in \autoref{tab:masses}, with the dynamical mass estimates from the accompanying work of \citet{exoALMA:Cristiano}. In addition, we also consider the estimates based on C$^{18}$O for the MAPS disks \citep{Zhang2021}, which we already showed in section \ref{sec:surface_density_everything}. The result of the comparison is shown in \autoref{fig:comparison_masses}.

The first thing to note is that our mass estimates are systematically lower than the dynamical masses, pointing strongly towards CO depletion as a likely explanation. In addition, it is reassuring that the sources where \citet{exoALMA:Cristiano} reports only upper limits on the disk mass (denoted as arrows in the figure) correspond to the lowest masses in the sample also with our methodology.

For what concerns the comparison with the optically thin derived mass, we note first that the strongest discrepancy is MWC480. This discrepancy is driven by radii smaller than 100 au, where the surface density reported by \citet{Zhang2021} increases steeply - outside this radius there is reasonable agreement with our results. The origin of this discrepancy is unclear. A possible explanation could be a high level of CO depletion spatially localized in the upper layers of the inner disk, explaining why other regions of the disk are in better agreement. It is worth noting that in the inner disk the rotation curve estimate (\citealt{Martire2024}, and see also \citealt{AndrewsRotationCurve} for a more flexible parametrization that gives similar results) is comparable to the results of \citet{Zhang2021}, possibly implying no CO depletion at those radii.

Even considering MWC 480, in the MAPS sample the $^{13}$CO mass is on average a factor 1.25 higher than the C$^{18}$O estimate. Such an agreement is remarkable considering our estimates are based on optically thick emission. For what concerns $^{12}$CO, when taking into account uncertainties there is also good agreement at face value, with an average of 1.3. Looking however at individual sources tells a different story, with GM Aur and HD163296 (which as mentioned in \ref{sec:compare_12_13} are among the largest discrepancies between $^{12}$CO and $^{13}$CO) being severely overestimated (but with large error bars) and MWC 480 that we already discussed. As mentioned in the previous section, for the first two the discrepancy is driven by the $^{12}$CO surface density reaching very high values in the inner ($\lesssim 100$ au) disk.

\begin{figure*}
    \centering
    \includegraphics[width=\textwidth]{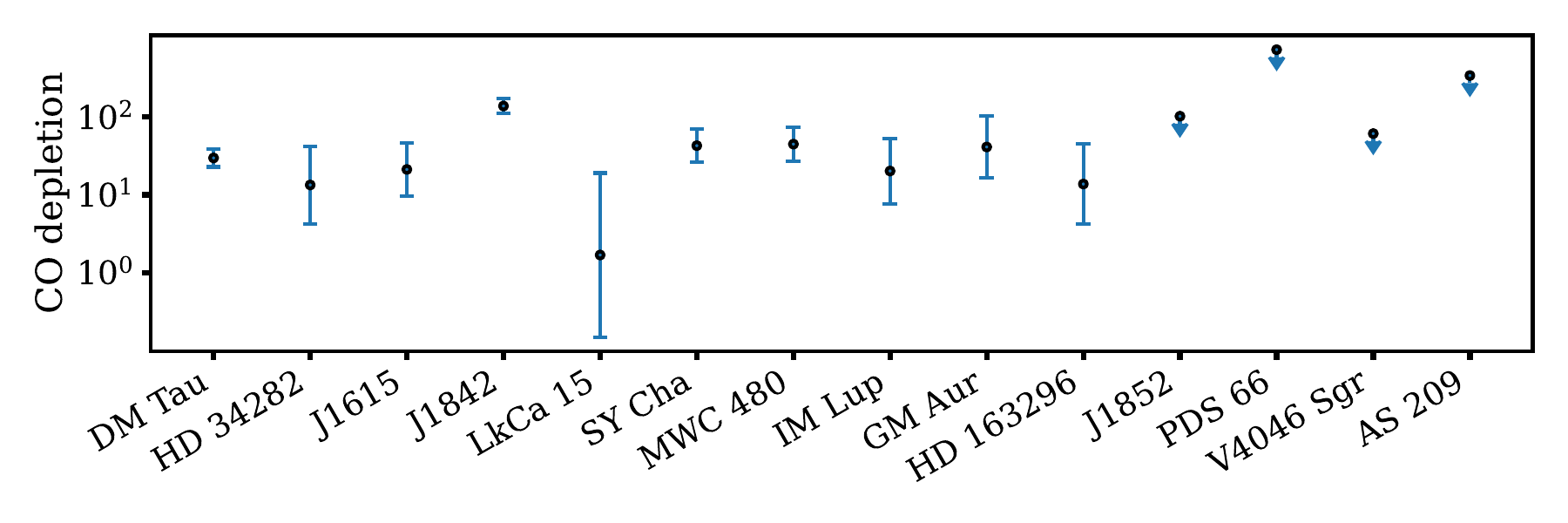}
    \caption{CO depletion factors needed with respect to the ISM value $x_\mathrm{CO}=10^{-4}$ to match the masses derived in this work from $^{13}$CO with the kinematical masses derived in \citet{exoALMA:Cristiano} and \citet{Martire2024}.}
    \label{fig:CO_depletion}
\end{figure*}

We now consider more in detail what levels of CO depletion are implied by our results. We show in \autoref{fig:CO_depletion} the ratio between the dynamical mass estimates and the $^{13}$CO masses (using this or the $^{12}$CO one is arbitrary since we showed they are statistically the same). Most of the datapoints cluster around a value of a few tens. Excluding the disks that have only upper limits from the rotation curve, the median CO depletion factor is $\sim$20 and the average (taking into account error bars) CO depletion factor is $\sim$ 50. This value is in line with what found recently by \citet{Teresa2024} using a similar methodology to what we employ here. More broadly, our results confirm the mounting evidence \citep[e.g.,][]{McClure2016,Miotello2017,BerginWilliams2017} that CO appears heavily depleted in proto-planetary disks. The sample we present here is currently the largest sample for which the mass is robustly constrained from the rotation curve.

The depletion of CO is probed observationally as part of this paper series also by another study \citep{exoALMA:Leon}. In that case, the emission lines used probe deeper into the disk midplane, and are therefore more sensitive to the levels of depletion in the midplane rather than in the disk upper layers. We refer to the paper for an extensive discussion of their findings.

\subsection{Caveats and future prospects}

As repeatedly stated our surface density estimates are extrapolations from the emitting height down to the midplane, and as such they require knowledge of the disk temperature profile. In this work we have employed a parametrization of the disk temperature motivated on previous works, but it is likely that real disks have a more complex temperature structure. Our error analysis has shown that errors in the emitting height dominate over errors in temperature, but we reached this conclusion using only the statistical errors on the temperature; it is considerably more complex to estimate the uncertainty due to the chosen temperature parametrization. In addition, while we showed that the isothermal model produces similar results to our best estimate when using the line brightness temperature, this is only a validation of the former against the latter, and in principle we cannot exclude that \textit{both} estimates could be off from the correct value. If in the future more flexible, or even non-parametric, temperature structures become available it would be important to assess their impact on our results.

Another aspect where our estimates could be improved consists in the determination of the emitting heights, since the two methods employed in exoALMA can sometimes have notable differences (see the appendix of \citealt{exoALMA:Maria}, and appendix \ref{sec:ds_dm}). As shown, our uncertainties are currently dominated by the emitting height determination, and improvement in the accuracy of this estimate is therefore an obvious point where to make progress. 

In this work we use the J=3-2 transition from the exoALMA sample and the J=2-1 transition from the MAPS sample for $^{12}$CO and $^{13}$CO. Those two samples currently constitute the best available observations, but we note that in the literature there are already estimates \citep{Law2022,Law2023, Law2024, Paneque-Carreno2023, Stapper2023} of the emitting height for additional sources in CO molecular emission, albeit at lower data quality. Only in some of these cases there are also temperature fits available, and the extraction procedure of the emitting height is not homogeneous across the literature - further work would be required before we can apply our method to these sources, but it should be possible. Considering there are ongoing ALMA Large Programs to survey gas emission in large samples (e.g. DECO and CHEER) and that they should be able to measure the emitting height at least for the brightest and largest disks, it is likely the observational sample our technique can be applied to will expand in the future. In addition, more isotopologues and transitions are observable with ALMA and in the future this work could be extended in this direction. In particular, multiple transitions would allow a more precise measurement of the vertical temperature profile, in a similar fashion to what was done in \citet{Pezzotta2025} for the measurement of the rotation curve, rather than using only two as we employ here. We also remark that it should be especially useful to consider high-J transitions that have a significantly different opacity from those we considered here. Different transitions of the same molecule, with an accurate measurement of the temperature profile, would also allow us to study how the abundance of the molecule changes at different heights, along similar lines to what we did here when comparing the surface density inferred from $^{12}$CO and $^{13}$CO to investigate the isotopic ratio.

In this work we have derived the CO depletion factor comparing our results with the dynamical mass. In the last few years, chemical methods have been devised \citep{Anderson2019,Anderson2022,Trapman2022} to infer the same factor using observations of the same disk in CO isotopologues and N$_2$H$^+$. A comparison between the chemical methods and the dynamical masses is presented as part of this paper series in \citet{exoALMA:Leon}. They find lower levels of depletion than what we report here, and suggest this may be because their method is probing deeper layers in the disk, where CO may be less depleted.

Finally, in this work we focused on the CO molecule. We note however that our model could be easily extended to other optically thick molecules. The only requirement is to have a simple model describing how the abundance of the molecule changes throughout the disk. If this is not available, thermochemical models remain the only way to link the observed emitting heights with the disk physical structure.

\section{Conclusions}
\label{sec:conclusions}

In this paper, we introduced a semi-analytical model linking the disk density and temperature structure with the emitting height of the CO molecule. The model depends on a free parameter describing photo-dissociation that we calibrated against thermochemical calculations.

We then applied our model to the MAPS and exoALMA disk samples where the CO emitting height has been measured. Our model turns these measurements into measurements of the CO column and therefore, assuming a CO abundance, into a disk surface density. By integrating this value we also provide measurements of the disk mass.

The key findings of our work are as follows:

\begin{itemize}
    \item In this work we provide a method to measure the disk surface density from optically \textit{thick}, rather than \textit{thin} as commonly done, emission. This comes at the cost of increased uncertainty. We find that the dominant source of error in our measurements is the error on the emitting height of the transition.
    \item In the sample we consider the temperature has been determined using two different, optically thick transitions. We show that when only one transition is available the use of a vertically isothermal model still yields satisfactory constraints on the disk surface density if using the line brightness temperature to set the disk temperature.
    \item Our model yields two different constraints for the $^{12}$CO and the $^{13}$CO molecule. In general they are in relatively good agreement assuming a standard isotopic ratio, which is a confirmation that our results are robust. The $^{13}$CO masses are systematically higher by a factor $\sim$2.5, in line with recent findings in other works of variations of the isotopic ratio in proto-planetary disks, but given the large uncertainties we are unable to further investigate this difference.
    \item We compared the masses we obtain with the dynamical fits to the disk rotation curve. Our values are in general lower, implying that the CO abundance needs to be reduced with respect to the standard ISM value we assumed. A median CO depletion of a factor $\sim$20 is needed to reconcile our mass measurements with the dynamical values, in line with other results showing that CO is depleted in proto-planetary disks.
\end{itemize}

\section{Acknowledgments}

We thank an anonymous referee for their comments which improved the clarity of the paper. This paper makes use of the following ALMA data: ADS/JAO.ALMA\#2021.1.01123.L. ALMA is a partnership of ESO (representing its member states), NSF (USA) and NINS (Japan), together with NRC (Canada), MOST and ASIAA (Taiwan), and KASI (Republic of Korea), in cooperation with the Republic of Chile. The Joint ALMA Observatory is operated by ESO, AUI/NRAO and NAOJ. The National Radio Astronomy Observatory is a facility of the National Science Foundation operated under cooperative agreement by Associated Universities, Inc. We thank the North American ALMA Science Center (NAASC) for their generous support including providing computing facilities and financial support for student attendance at workshops and publications.

JB acknowledges support from NASA XRP grant No. 80NSSC23K1312. MB, DF, JS have received funding from the European Research Council (ERC) under the European Union’s Horizon 2020 research and innovation programme (PROTOPLANETS, grant agreement No. 101002188). Computations by JS have been performed on the `Mesocentre SIGAMM' machine, hosted by Observatoire de la Cote d’Azur. PC acknowledges support by the Italian Ministero dell'Istruzione, Universit\`a e Ricerca through the grant Progetti Premiali 2012 – iALMA (CUP C52I13000140001) and by the ANID BASAL project FB210003. SF is funded by the European Union (ERC, UNVEIL, 101076613), and acknowledges financial contribution from PRIN-MUR 2022YP5ACE. MF is supported by a Grant-in-Aid from the Japan Society for the Promotion of Science (KAKENHI: No. JP22H01274). JDI acknowledges support from an STFC Ernest Rutherford Fellowship (ST/W004119/1) and a University Academic Fellowship from the University of Leeds. Support for AFI was provided by NASA through the NASA Hubble Fellowship grant No. HST-HF2-51532.001-A awarded by the Space Telescope Science Institute, which is operated by the Association of Universities for Research in Astronomy, Inc., for NASA, under contract NAS5-26555. CL has received funding from the European Union's Horizon 2020 research and innovation program under the Marie Sklodowska-Curie grant agreement No. 823823 (DUSTBUSTERS) and by the UK Science and Technology research Council (STFC) via the consolidated grant ST/W000997/1. CP acknowledges Australian Research Council funding  via FT170100040, DP18010423, DP220103767, and DP240103290. DP acknowledges Australian Research Council funding via DP18010423, DP220103767, and DP240103290. GR acknowledges funding from the Fondazione Cariplo, grant no. 2022-1217, and the European Research Council (ERC) under the European Union's Horizon Europe Research \& Innovation Programme under grant agreement no. 101039651 (DiscEvol). H-WY acknowledges support from National Science and Technology Council (NSTC) in Taiwan through grant NSTC 113-2112-M-001-035- and from the Academia Sinica Career Development Award (AS-CDA-111-M03). GWF acknowledges support from the European Research Council (ERC) under the European Union Horizon 2020 research and innovation program (Grant agreement no. 815559 (MHDiscs)). GWF was granted access to the HPC resources of IDRIS under the allocation A0120402231 made by GENCI. Support for BZ was provided by The Brinson Foundation. CH acknowledges support from NSF AAG grant No. 2407679. GL has received funding from the European Union's Horizon 2020 research and innovation program under the Marie Sklodowska-Curie grant agreement No. 823823 (DUSTBUSTERS). TCY acknowledges support by Grant-in-Aid for JSPS Fellows JP23KJ1008. AJW has received funding from the European Union's Horizon 2020 research and innovation programme under the Marie Skłodowska-Curie grant agreement No 101104656. Views and opinions expressed by ERC-funded scientists are however those of the author(s) only and do not necessarily reflect those of the European Union or the European Research Council. Neither the European Union nor the granting authority can be held responsible for them. T.PC thanks the Heising-Simons Foundation for their funding through the 51 Pegasi b Postdoctoral Fellowship. The National Radio Astronomy Observatory and Green Bank Observatory are facilities of the U.S. National Science Foundation operated under cooperative agreement by Associated Universities, Inc.


%

\vspace{5mm}
\facilities{ALMA}


\software{DALI}



\appendix

\section{Comparison between \disksurf and \discminer emitting heights}
\label{sec:ds_dm}

\begin{figure}
    \centering
    \includegraphics[width=\textwidth]{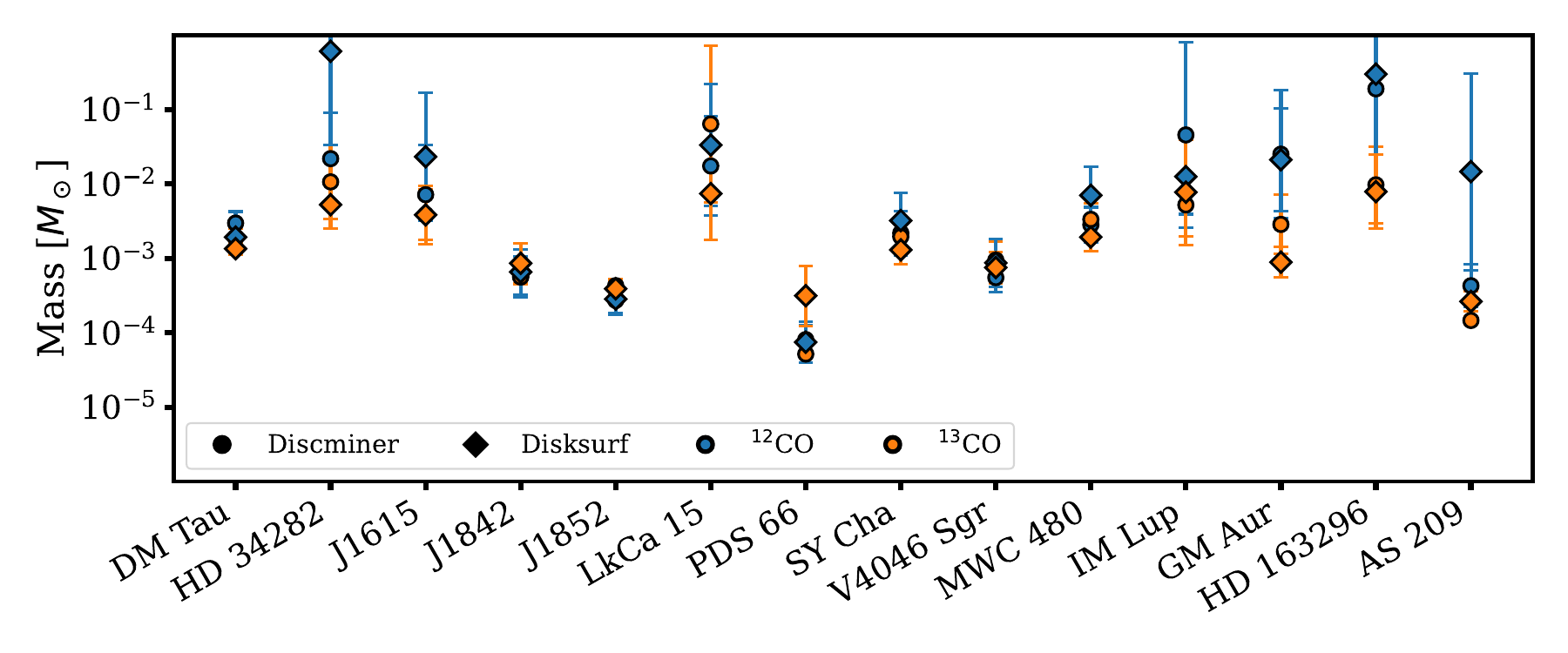}
    \caption{Comparison between the disk masses obtained using the emitting heights from \disksurf and those from \discminer.}
    \label{fig:ds_dm}
\end{figure}

We show in \autoref{fig:ds_dm} a comparison between the disk masses obtained using the emitting heights from \disksurf and those from \discminer. The two estimates agree across the sample, with an average ratio of 0.9 for $^{12}$CO and of 1.8 for $^{13}$CO, meaning there is no particular bias between the two methods, but the estimates can vary significantly source by source. Indeed, the standard deviation of the mass ratio is 0.6 dex for $^{12}$CO and of 0.4 dex for $^{13}$CO. The plot confirms visually that the discrepancy can be very significant for $^{12}$CO (particularly recognizable, as expected from the comparison shown in \citealt{exoALMA:Maria}, are HD 34282 and J1615), although when this happens the error bars are also large.

Considering that the disk mass uncertainties we report in this paper are comparable to the standard deviation we reported above, our results will not significantly change due to the choice of the way in which the emitting heights are retrieved. On the other hand, the differences that do emerge also show that there is still work to do in finding and developing the best method to retrieve emission heights.

\section{Isothermal approach using $^{12}$CO and $^{13}$CO at the same time}
\label{app:system_two_equations}

We describe in the main text in section \ref{sec:comparei} the results of the isothermal model when using the temperature at the emitting height. In this appendix we describe the results when using the other approach we introduced in section \ref{sec:inverse_isothermal}, namely using two different isotopologues at the same time to jointly derive the scale-height $H$ and the disk surface density.

\begin{figure}
    \centering
    \includegraphics[width=0.48\columnwidth]{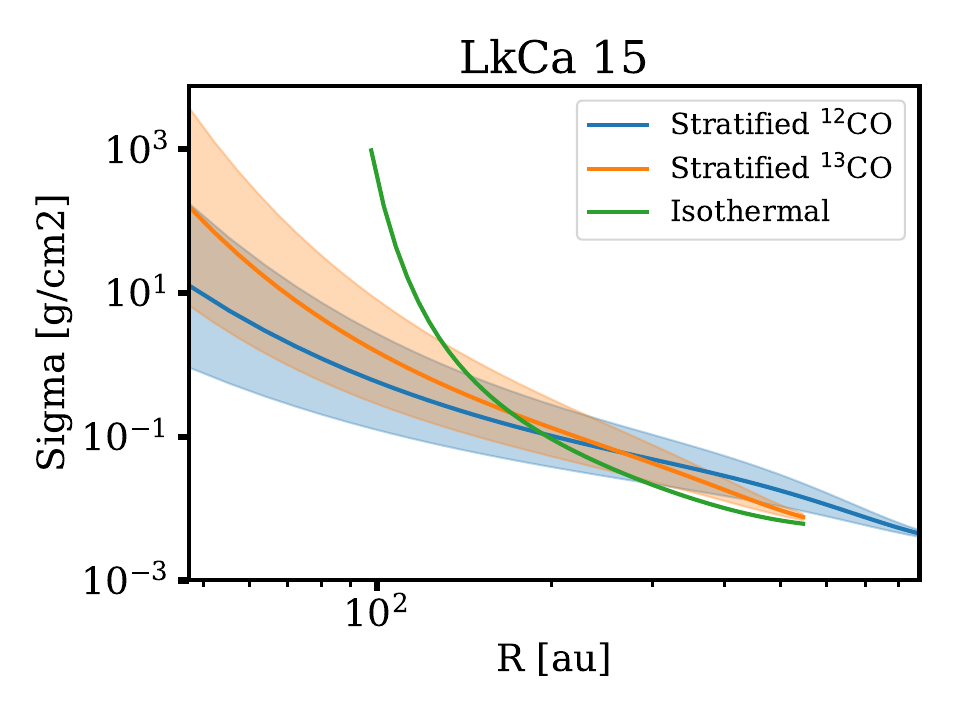}
    \includegraphics[width=0.48\columnwidth]{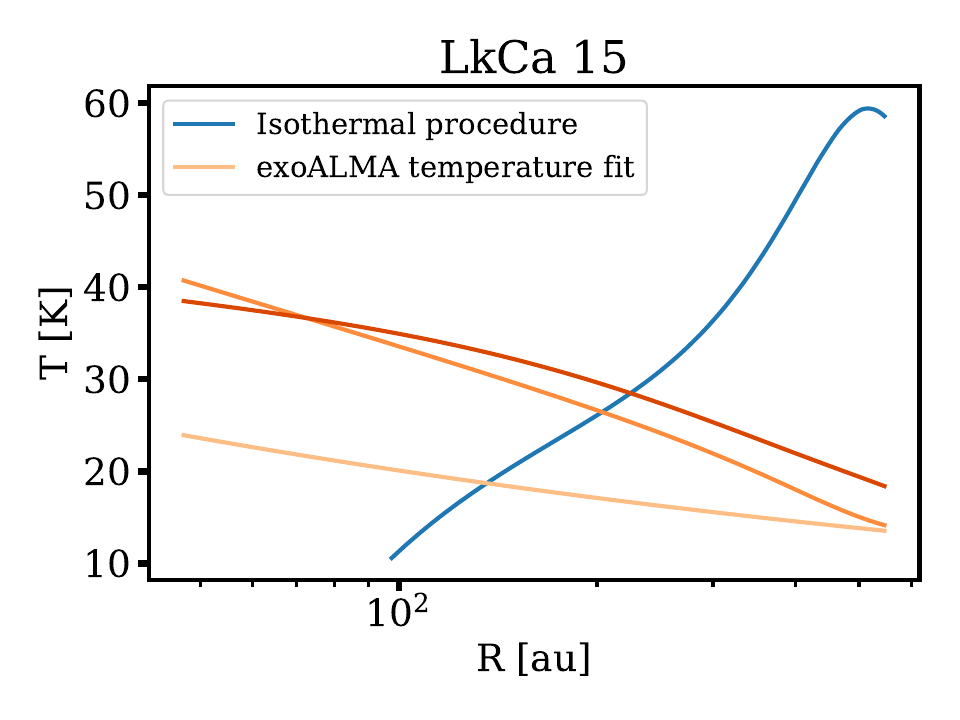}
    \includegraphics[width=0.48\columnwidth]{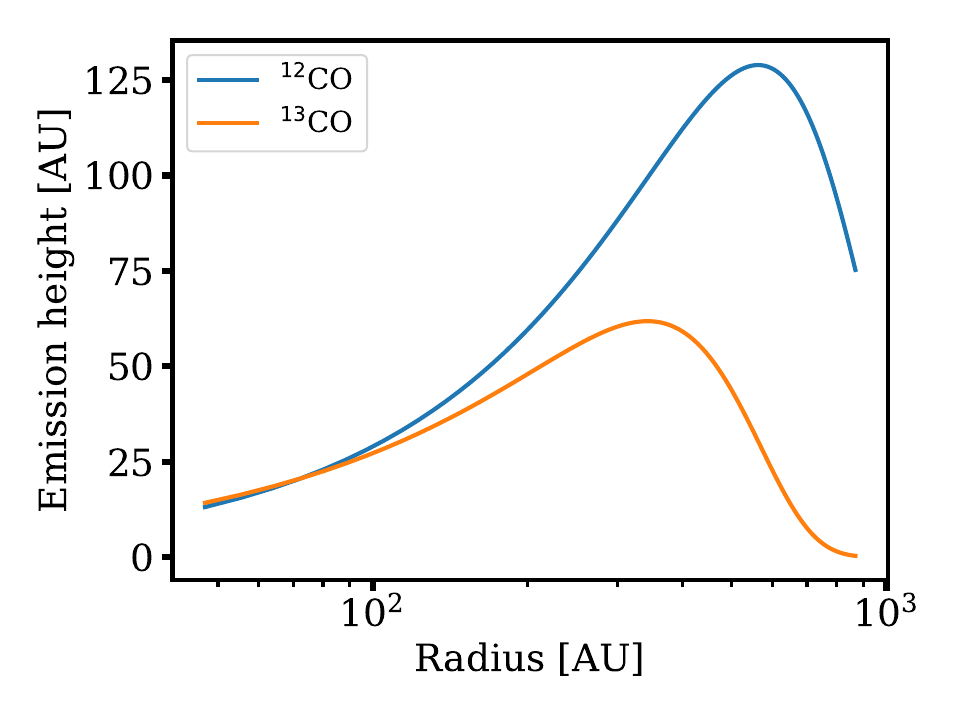}
    \caption{\textbf{Left panel:} surface density $\Sigma$ in LkCa 15 of the stratified models in comparison with the isothermal model when solving for the scale-height using both the $^{12}$CO and $^{13}$CO emitting height at the same time. \textbf{Right panel: } comparison between the temperatures returned by the solution shown in the left panel and the temperature fit reported in \citet{exoALMA:Maria} (at three different layers: the extrapolation to the midplane and the values at the emitting heights of $^{12}$CO and $^{13}$CO). \textbf{Bottom panel:} comparison between the emitting heights of $^{12}$CO and $^{13}$CO, showing that in the inner disk the former is smaller than the latter.}
    \label{fig:results_lkca15_iso}
\end{figure}

For the example case of LkCa 15, we show the results of the isothermal model in comparison with the stratified one in the first panel of \autoref{fig:results_lkca15_iso}. While in the outer disk the retrieved surface density is broadly comparable to the stratified case, we note that in the inner disk the isothermal profile is significantly denser. To illustrate why, we need to consider the second panel, where we show a comparison between the temperature returned by the isothermal model and the 2D temperature fit \citep{exoALMA:Maria}. Since in the latter case the temperature is a function of the vertical coordinate, we show the temperatures at three different layers: the extrapolation to the midplane and the values at the emitting heights of $^{12}$CO and $^{13}$CO. It can be seen how the temperature returned by the isothermal model becomes unphysically low in the inner disk. Eventually, for radii smaller than $\approx$ 100 au the isothermal profile is not defined and no value for the surface density and temperature can be found that solves the system of equations defined in section \ref{sec:inverse_isothermal}.

We attribute the behavior in the inner disk to the fact that the emitting heights of $^{12}$CO and $^{13}$CO become similar within 100 au, and within $\sim$60 au the emitting height of $^{13}$CO is higher than that of $^{12}$CO, as can be seen in the third panel of the figure. In our model by construction $^{13}$CO must come from a deeper layer than $^{12}$CO, explaining why no solution can be found inside $\sim$60 au. At the radii where two heights are similar, \autoref{eq:H_isothermal} implies that the solution must have a small $H$, so that both emitting heights are far from the center of the Gaussian, in the region where the density rapidly drops. Such a high $^{13}$CO emitting surface also explains why even for the vertically stratified case the surface density measured from $^{13}$CO is significantly higher in the inner disk than when measuring it from $^{12}$CO. Considering that $^{12}$CO is significantly more abundant than $^{13}$CO, this behaviour is clearly unphysical and it is likely driven by errors in the retrieval of the emitting height, in particular for $^{13}$CO. Because we use a parameterised surface in \discminer, it is possible that a more flexible parametrization would alleviate the issue. 

This behavior is not unique of LkCa 15 - we find no solution for the isothermal model also in the inner disk of V4046 Sgr. In addition, MWC 480 and SY Cha (in this case in the outer disk) are also severely affected by the issue. Formally, we are still able to find solutions because the $^{13}$CO emitting height is smaller than the $^{12}$CO one, but because they become very similar the isothermal surface density estimate becomes orders of magnitude higher than the stratified case.  

Even when excluding such extreme cases, we find that using the two emitting height measurements at the same time often performs significantly less well than when using the brightness temperature at the emission surface; in some cases it yields disk masses that can differ up to one order of magnitude from either of the stratified estimates. We attribute this failure to the fact that ultimately $^{12}$CO and $^{13}$CO come from layers with different temperature. Because the model we discuss in this appendix uses their relative ratio to set the scale-height, it is more sensitive to the vertical variations of the temperature than the models using a single emission line to estimate the surface density. Because of these reasons, we will not consider further using both emitting heights at the same time.

\section{Density profile comparison between isothermal and stratified models}
\label{app:extrapolation}

\begin{figure}
\centering
\includegraphics[width=0.5\columnwidth]{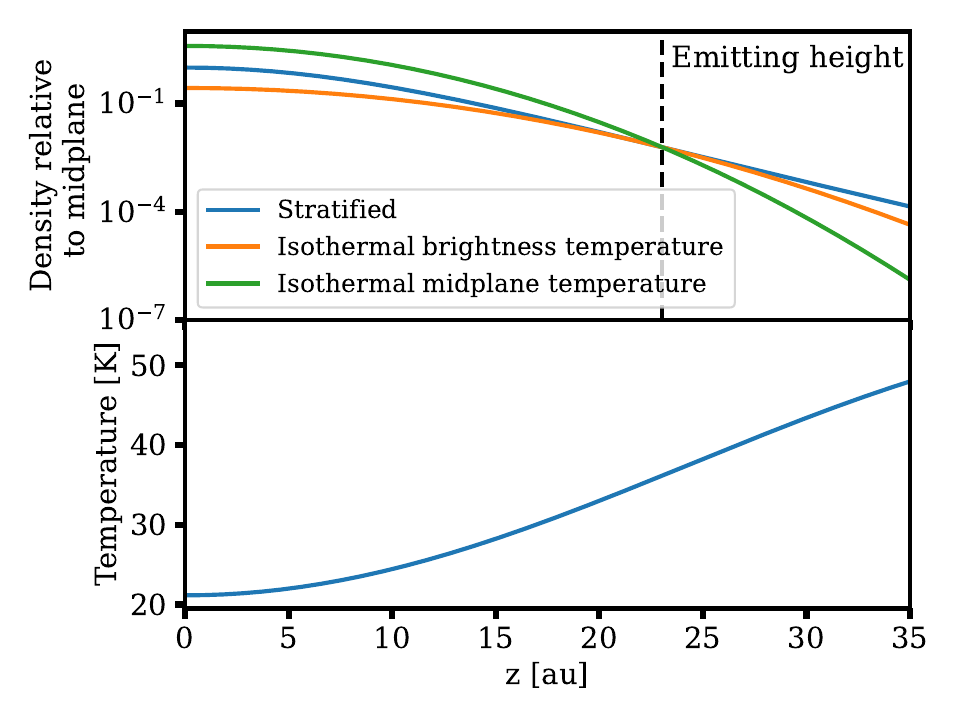}
\caption{\textbf{Top panel:} density profiles in LkCa 15 at 80 au. We show the density profiles for the stratified case and for the two isothermal cases described in section \ref{sec:comparei}. The stratified profile is normalized to 1 in the midplane, and the other two profiles are normalized so that at the $^{12}$CO emitting height (marked by the dashed vertical line) they intersect with the stratified one. \textbf{Bottom panel}: temperature profile derived from the \citet{exoALMA:Maria} fit.}
\label{fig:extrapolation}
\end{figure}

Our method can essentially be understood as an extrapolation from the conditions at the emitting height to the midplane. The extrapolation assumes hydrostatic equilibrium and as such relies on knowledge of the temperature structure. To visualize in practice how the extrapolation happens, we show in the upper panel of \autoref{fig:extrapolation} the vertical density profile for LkCa 15 at a reference radius of 80 au, comparing the stratified model where the temperature varies with the vertical coordinate, as shown in the bottom panel, with the two isothermal models described in the text, one where we set the temperature to the value at the emitting height and one where we set the temperature to the midplane value. We also mark the $^{12}$CO emitting height with the dashed vertical line.

We normalize the profiles so that they all pass by the same point at the emitting height, with a value set by the stratified case with respect to the midplane (i.e., for the stratified case the density in the midplane is normalized to 1). This is justified by the fact that, to first order, the models constrain the density to be similar (but not mathematically equal) at the emitting height, since they all require the line emission to become optically thick, and the column integrals are dominated by the region close to the emitting height. The plot shows visually that the value returned by the extrapolation for the midplane density depends on the steepness of the density profile; for example, the midplane temperature is the lowest, implying a steeper profile that returns the \textit{highest} midplane density. The opposite happens when using the emitting height temperature, which therefore produces the \textit{lowest} midplane density; the stratified case falls somewhere in between. We find this ordering to apply to most cases, though there are exceptions. We interpret this as due to the fact that the optical depth depends on the whole vertical integral up to infinity. The steepness of the density profile has therefore an impact in the density retrieved at the emitting height: for example, both isothermal cases fall off more steeply than the stratified case since the temperature will still increase in the upper layers, yielding for both a slightly higher density at the emitting point. This further exacerbates the overestimate for the midplane temperature case, while it partially compensates the underestimate for the emitting height temperature one, explaining why the latter generally fares better than the former. In some rare cases, we even find that the latter effect dominates and that the emitting height temperature model overestimates the stratified value. In practice, we note that the exact magnitude of the under/over estimate depends on the exact temperature profile, and even qualitatively there can be a large variety of cases: for example, the emitting height could be located at a height where the temperature profile is close to the midplane value, still increasing (as it is here) or where it has plateaued to the surface value. We can easily imagine that in the first case (temperature at the emitting height close to the midplane value) using the midplane temperature should give an estimate very close to the isothermal value.

The arguments above justify why in general the midplane temperature gives a worse estimate than the emitting height temperature, but we remark that the truth of the statement depends somehow on the temperature profile. Bearing in mind that both isothermal models are approximations, we consider this result as empirical and it can only be justified \textit{a posteriori}, as we have done here.


\bibliography{emission_height}{}
\bibliographystyle{aasjournal}



\end{document}

%% file: table_masses.tex
\begin{tabular}{lrrrr}
\toprule
{} & $^{12}$CO mass [$M_\odot$] & Uncertainty [dex] & $^{13}$CO mass [$M_\odot$] & Uncertainty [dex] \\
\midrule
DM Tau    &       $3.0 \times 10^{-3}$ &               0.2 &       $1.9 \times 10^{-3}$ &               0.1 \\
HD 34282  &       $2.2 \times 10^{-2}$ &               0.6 &       $1.1 \times 10^{-2}$ &               0.5 \\
J1615     &       $7.2 \times 10^{-3}$ &               0.7 &       $3.9 \times 10^{-3}$ &               0.3 \\
J1842     &       $5.7 \times 10^{-4}$ &               0.3 &       $5.6 \times 10^{-4}$ &               0.1 \\
J1852     &       $2.7 \times 10^{-4}$ &               0.2 &       $4.3 \times 10^{-4}$ &               0.1 \\
LkCa 15   &       $1.7 \times 10^{-2}$ &               0.7 &       $6.4 \times 10^{-2}$ &               1.1 \\
PDS 66    &       $8.1 \times 10^{-5}$ &               0.2 &       $5.2 \times 10^{-5}$ &               0.0 \\
SY Cha    &       $2.2 \times 10^{-3}$ &               0.3 &       $2.0 \times 10^{-3}$ &               0.2 \\
V4046 Sgr &       $5.5 \times 10^{-4}$ &               0.2 &       $9.5 \times 10^{-4}$ &               0.3 \\
MWC 480   &       $2.8 \times 10^{-3}$ &               0.2 &       $3.3 \times 10^{-3}$ &               0.2 \\
IM Lup    &       $4.6 \times 10^{-2}$ &               1.2 &       $5.2 \times 10^{-3}$ &               0.4 \\
GM Aur    &       $2.5 \times 10^{-2}$ &               0.9 &       $2.9 \times 10^{-3}$ &               0.4 \\
HD 163296 &       $1.9 \times 10^{-1}$ &               1.3 &       $9.7 \times 10^{-3}$ &               0.5 \\
AS 209    &       $4.3 \times 10^{-4}$ &               0.3 &       $1.5 \times 10^{-4}$ &               0.1 \\
\bottomrule
\end{tabular}